\documentclass[namedreferences,hyperref,optionalrh,solaromanenum]{spr-sola}
\usepackage{fix-cm}
\usepackage{graphicx}                    % For eps figures, newer & more powerfull
\usepackage{rotating}
\usepackage{color}                       % For color text: \color command
\usepackage[T1]{fontenc}
\usepackage{amsmath}
\chardef\us=`\_

\begin{document}
\begin{frontmatter}
\title{Bo\v{s}kovi\'{c}'s spherical trigonometric solution for determining the axis and rate of solar rotation by observing sunspots in 1777}
\author[addressref={aff1},corref,email={mhusak@geof.hr}]{\inits{M.}\fnm{Mirko}~\lnm{Husak}\orcid{0000-0002-2814-5119}}\sep
\author[addressref={aff2}       ,%__email={romanb@geof.hr}
]{\inits{R.}\fnm{Roman}~\lnm{Braj\v{s}a}\orcid{orcid.org/0009-0008-3048-8138}}\sep
\author[addressref={aff3}       ,%__email={dspoljar@geof.hr}
]{\inits{D.}\fnm{Dragan}~\lnm{\v{S}poljari\'c}\orcid{orcid.org/0000-0001-7427-044X}}\sep
\author[addressref={aff4}       ,%__email={dkrajnovic@aip.de}
]{\inits{D.}\fnm{Davor}~\lnm{Krajnovi\'c}\orcid{0000-0002-0470-6540}}\sep
\author[addressref={aff2}       ,%__email={rdomagoj@geof.hr}
]{\inits{D.}\fnm{Domagoj}~\lnm{Ru\v{z}djak}\orcid{0000-0002-4861-0225}}\sep
\author[addressref={aff6}       ,%__email={ivica.skokic@gmail.com}
]{\inits{I.}\fnm{Ivica}~\lnm{Skoki\'c}\orcid{0000-0002-5486-4192}}\sep
\author[addressref={aff5}       ,%__email={drosa@zvjezdarnica.hr}
]{\inits{D.}\fnm{Dragan}~\lnm{Ro\v{s}a}\orcid{}}\sep
\author[addressref={aff5}       ,%__email={dhrzina@zvjezdarnica.hr}
]{\inits{D.}\fnm{Damir}~\lnm{Hr\v{z}ina}\orcid{}}\sep

\runningauthor{M. Husak et al.}
\runningtitle{Bo\v{s}kovi\'{c}'s spherical trigonometric solution for determining the axis and rate of solar rotation...}% by observing sunspots in 1777}
\address[id={aff1}]{Trako\v{s}\'canska 20, 42000 Vara\v{z}din}
\address[id={aff2}]{Hvar Observatory, Faculty of Geodesy, University of Zagreb, Ka\v{c}i\'ceva 26, 10000 Zagreb, Croatia}
\address[id={aff3}]{Faculty of Geodesy, University of Zagreb, Ka\v{c}i\'ceva 26, 10000 Zagreb, Croatia}
\address[id={aff4}]{Leibniz-Institut f\"ur Astrophysik Potsdam (AIP), An der Sternwarte 16, 14482 Potsdam, Germany}
\address[id={aff6}]{Astronomical Society "Anonymus", Bra\'ce Radi\'ca 34, 31550 Valpovo, Croatia}
\address[id={aff5}]{Zagreb Astronomical Observatory, Opati\v{c}ka 22, 10000 Zagreb, Croatia}
\begin{abstract}
In 1777 Ru\dj er Bo\v{s}kovi\'{c} observed and measured sunspot positions to determine solar rotation elements. In 1785, among other methods, he described a trigonometric spherical solution for determination of the position of the axis and rate of solar rotation using three sunspot positions, but without equations. For the first time, we derive equations applicable for modern computers for calculating solar rotation elements as Bo\v{s}kovi\'{c} described. We recalculated Bo\v{s}kovi\'{c}'s original example using his measurements of sunspot positions from 1777 using the equations developed here and confirmed his results from 1785. Bo\v{s}kovi\'{c}'s methodology of arithmetic means determines $i$, $\Omega$, and sidereal period $T'$ separately, the planar trigonometric solution determines $i$ and $\Omega$ together, but his spherical trigonometric solution calculates $i$, $\Omega$, and sidereal period $T'$ in a single procedure. 
\end{abstract}
\end{frontmatter}
\begin{article}
\section{Introduction}
\label{s:Introduction} 
The application of astronomical telescopes in 1609 enabled precise measurements of phenomena on the apparent solar disk, while invention of logarithms in the early 17th century made demanding scientific calculations easier. John Napier\footnote{Scott, J. Frederick (2024, March 31). John Napier. Encyclopedia Britannica. https://www.britannica.com/biography/John-Napier} (1550—-1617) and Joost Bürgi\footnote{Britannica, T. Editors of Encyclopaedia (2024, February 24). Joost Bürgi. Encyclopedia Britannica. https://www.britannica.com/biography/Joost-Burgi} (1552—-1632) invented logarithms independently (\citealp{1614mlcd.book.....N}; \citealp{1620agp..book.....B}). Using logarithms, multiplication becomes addition: $\log (m \cdot n) = \log m + \log n$. This type of calculation is often used in astronomy.  % The logarithms enabled easier demanding scientific calculations.

Henry Briggs\footnote{Britannica, T. Editors of Encyclopaedia (2024, February 19). Henry Briggs. Encyclopedia Britannica. https://www.britannica.com/biography/Henry-Briggs} (1561—1630) in collaboration with Napier made logarithm tables with base $10$ (today we call it Common or Brigg's logarithms) in \textit{Logarithmorum Chilias Prima} \citep{1617lcp..book.....B}, \textit{Arithmetica Logarithmica} \citep{1624allc.book.....B}, and application of logarithms in trigonometry \textit{Trigonometria Britannica} \citep{1633trbr.book.....B}. 

Galileo Galilei was among the first who applied a telescope in astronomy in 1609. He observed the solar disk with a telescope in 1612 and he noticed the sunspots on the apparent solar disk, visible for 14 days, and again after about 30 days. He came to the conclusion that Sun rotates with the period about 30 days \citep{1613idim.book.....G}. 

Christoph \cite{1630rour.book.....S} was the first one who noticed the faster solar rotation of sunspots in the equatorial region than at the higher solar latitudes. Today, he is accepted as the discoverer of the differential solar rotation.  Observations of Christoph Scheiner were researched in \cite{2006SoPh..234..379C}. Much later, the solar differential rotation was precisely measured.

\cite{2020LRSP...17....1A} reviewed historical sunspot records, in pre-telescopic (naked-eye) and in telescopic period. There are drawings of sunspots on the solar disk of many researches such as Thomas Harriot in 1610, Galileo Galilei in 1611, Christoph Scheiner in 1612, Johann Caspar Staudacher in 1749 to 1796, Barnaba Oriani in 1778 to 1779, and many others, but only some of them determined, and few of them calculated solar rotation elements such as J. D. Cassini in 1678, J. Cassini in 1746, La Lande and Delambre in 1775, and Ru\dj er Bo\v{s}kovi\'{c} in 1777 \citep[Table 7]{2023SoPh..298..122H}. 

During the Maunder minimum (1645 - 1715) research of solar activity was challenging because there were fewer sunspots present on the Sun (\citealp{1976Sci...192.1189E} and \citealp{2006SoPh..234..379C}). Solar rotation in the 17th century was researched by \cite{1977Sci...198..824E}, \cite{2006SoPh..234..379C}, \cite{2022SoPh..297..132S}, and \cite{1982QJRAS..23..213Y}, and solar differential rotation in the 18th century \citep{2012A&A...543A...7A}.
Ru\dj er Bo\v{s}kovi\'{c} was the one who observed and measured sunspot positions on the apparent solar disk and he calculated solar rotation elements using numeric measurements.
\begin{figure}[ht]
    \centering
    \includegraphics[width=\textwidth]{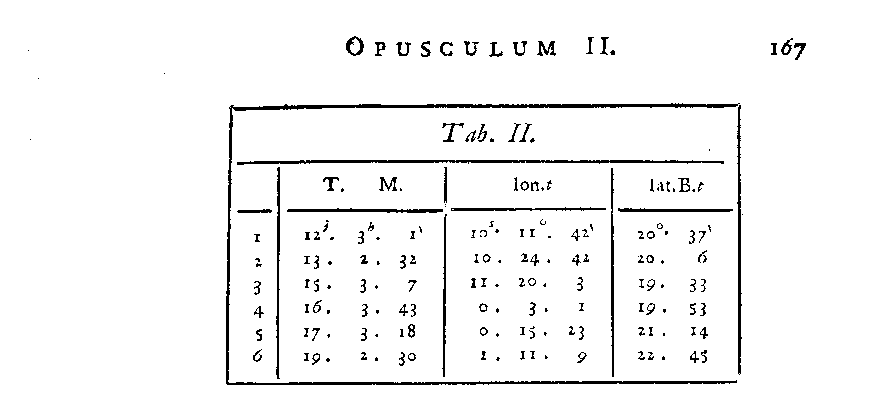}
    \caption{Positions of the sunspot: mean solar time $T.M.$ and ecliptic coordinates $lon.t$ and $lat.B.t$ of the first sunspot, which \cite{1785BoscovichII} observed and measured its positions on the apparent solar disk with astronomical telescope in 1777, and then he calculated $T.M.$, $lon.t$, and $lat.B.t$ from the measurements using trigonometry and logarithms in $Tab. II.$, p167.}
    \label{fig:tabII}
\end{figure}

\begin{figure}[]
    \centering
    \includegraphics[width=\textwidth]{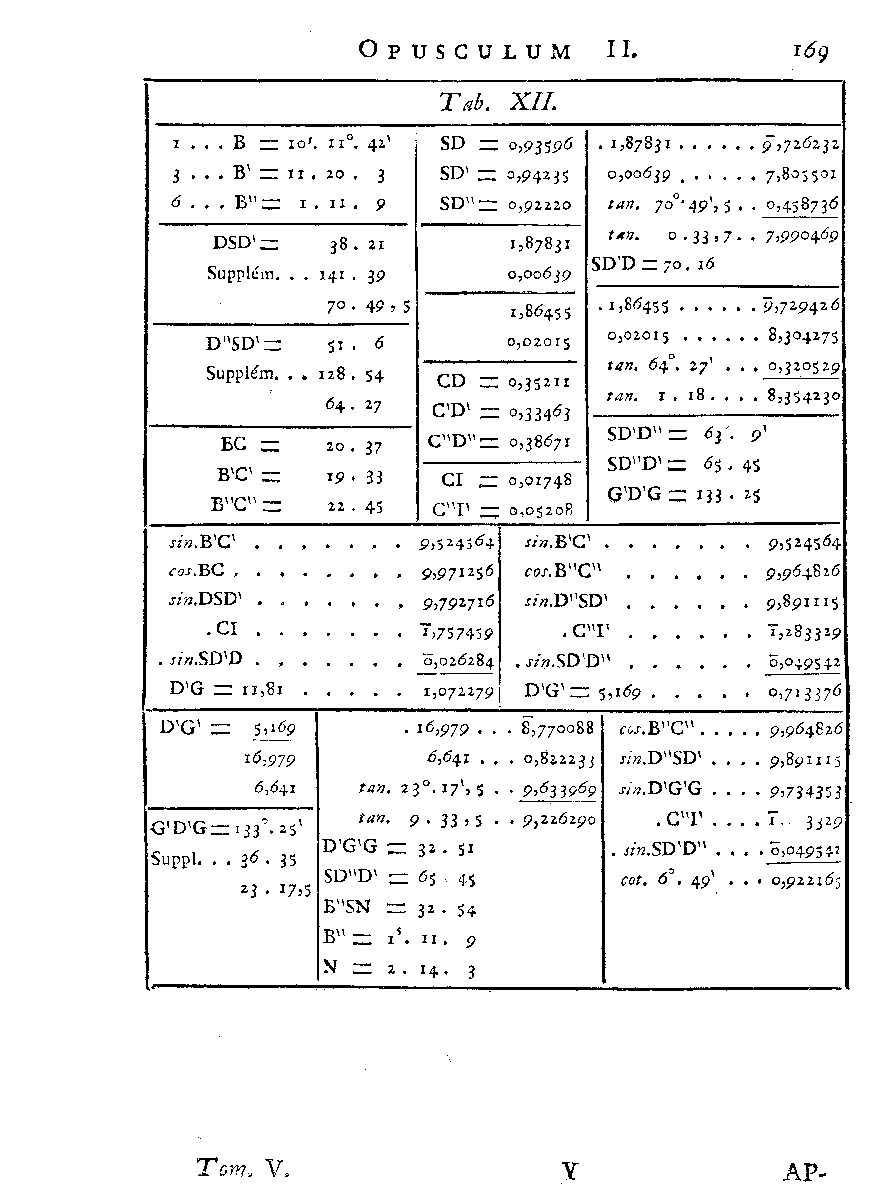}
    \caption{Planar trigonometric solution: logarithmic calculation of $\Omega$ and $i$, using three sunspot positions 1, 3, and 6 from \textit{Tab. II.} (Figure~\ref{fig:tabII}): $N=2^S 14^\circ03'=74^\circ03'=\Omega$ and $i=6^\circ49'$  \citep[\textit{Tab. XII.}, p169]{1785BoscovichII}.}
    \label{fig:TabXII}
\end{figure}

\begin{figure}[]
    \centering
    \includegraphics[width=\textwidth]{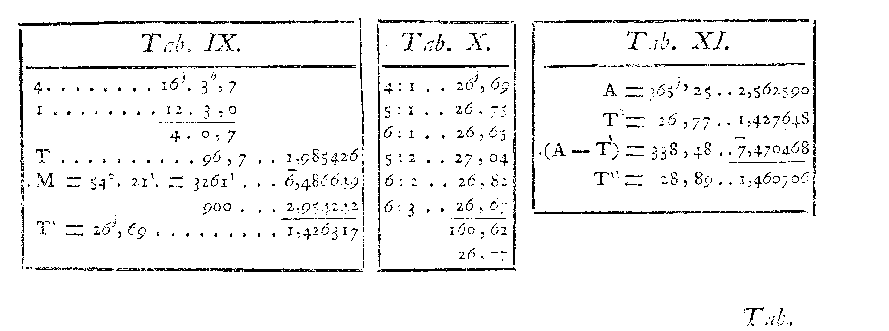}
    \caption{Solar rotation periods calculated from six sunspot pairs of positions of one sunspot: sidereal $T'=26.77$ days and synodic $T''=28.89$ days, calculated from $T'$ \citep[\textit{Tab. IX.}, \textit{Tab. X.}, and \textit{Tab. XI.}, p168]{1785BoscovichII}.}
    \label{fig:tabIXXXI}
\end{figure}

Ru\dj er Bo\v{s}kovi\'{c} used astronomical telescope for observations and these new mathematical methods of application logarithms and its application in trigonometry on his works in astronomy \citep{boscovich1785opera}.
Ru\dj er Bo\v{s}kovi\'{c} observed and measured sunspot positions on the solar disk and then he  calculated sunspot positions in ecliptic coordinate system using trigonometry and logarithms (Figure~\ref{fig:tabII}). Then, using the sunspot positions, he calculated solar rotation elements: the longitude of the ascending node $\Omega$, the solar equator inclination $i$ (Figure~\ref{fig:TabXII}), the solar rotation periods: the sidereal $T'$ and the synodic one $T''$ (Figure~\ref{fig:tabIXXXI}).

Independently, Richard Christopher \cite{1863oss..book.....C} and Friederich Wilhelm Gustav \cite{1874bsza.book.....S} observed and measured sunspot positions and then independently determined the solar rotation elements $\Omega$ and $i$. They confirmed solar differential rotation, lower angular velocity $\omega$ at higher heliographic latitudes $b$. %Angular velocity $\omega$ is inversely proportional to rotational period $T$. 
Carrington determined the mean synodic rotation period of sunspots of $27.2753$ days, which we call \textit{Carrington rotations} after him.

\subsection{Solar rotation elements}
\begin{figure}[h]
    \centering
    \includegraphics[width=\columnwidth]{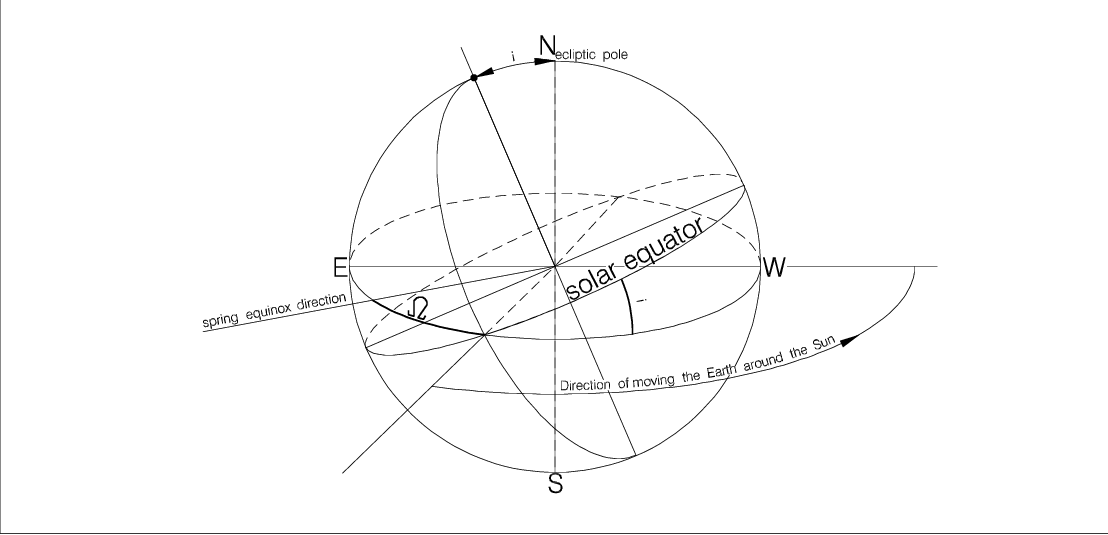}
    \caption{The Carrington solar rotation elements: $\Omega$ the longitude of the ascending node, and $i$ the inclination of the solar equator to the ecliptic.}% \citep{2022arXiv220311745H}.}
    \label{fig:SolarRotationElements}
\end{figure}

Solar rotation is defined with the period $T$, and the position of the solar rotation axis in space, the longitude of the ascending node $\Omega$ and the inclination of the solar equator $i$, e. g., \cite{2002tsai.book.....S} (Figure~\ref{fig:SolarRotationElements}). Today we use solar differential law $\omega(b)=A+B \cdot \sin ^2 b$, where $\omega(b)$ is angular velocity at heliographic latitude $b$, $A$ and $B$ we usually determine using $L_2$ (gaussian) least square fitting method (LSQ). Sidereal period we determine as $T=1/\omega(b)$.

%, many researchers which i
In Table 1 \cite{1978A&A....62..165W} presented determinations of $\Omega$ and $i$ since the application of the telescope in astronomy in 1609. %We widen determinations to the nowadays where we
We expanded the table with recent measurements and included also Bo\v{s}kovi\'{c}'s results \citep[Table 7]{2023SoPh..298..122H}.

\subsection{Solar rotation elements determinations by Ru\dj er Bo\v{s}kovi\'{c}}

Ru\dj er Bo\v{s}kovi\'{c} described his methods for determination of the sunspot positions, position of solar rotation axis and solar rotation rate by observing sunspots in his so-called dissertation \textit{De maculis solaribus} \citep{1736dmse.book.....B}. In 1777 he observed sunspots and measured their positions on the apparent solar disk using his own methods. In 1785, in the chapter \textit{Opuscule II} in French\footnote{\textit{Sur les éléments de la rotation du soleil sur son axe déterminés par l'observation de ses taches.}}, in 5th book of five-book compendium \textit{Opera pertinentia ad opticam et astronomiam} he published this complete astronomical and scientific experiment \citep{boscovich1785opera}. 

\textit{Opuscule II} includes description of his methods with drawings and equations, instruments he used, measurements of his observations\footnote{\textit{Appendice. Journal des observations de plusieurs taches du soleil faites à Noslon près de Sens chez S. E. M. le cardinal de Luynes l'année 1777.}} in 1777, detailed description of his method, and instructions for calculations. He calculated the results using trigonometry and logarithms, the results comprise of: sunspot positions of the first sunspot and solar rotation elements: $\Omega$ ecliptic longitude of the ascending node, $i$ solar equator's inclination, and solar rotation periods, sidereal $T'$ and synodic $T''$. He presented the results in twelve tables \textit{Tab. I.}~-\textit{Tab. XII.}  \citep{1785BoscovichII}. Solar rotation elements are presented in Figure~\ref{fig:SolarRotationElements}. 

\cite{2022arXiv220311745H} described the problem of Bo\v{s}kovi\'{c}'s determination of solar rotation elements using sunspot positions on the apparent solar disk \citep[\textit{Opuscule II}]{1785BoscovichII}. \cite{2022arXiv220302289H} repeat Bo\v{s}kovi\'{c}'s original logarithmic calculations of solar rotation elements $\Omega$, $i$, $T'$, and $T''$. Later we modernized the original equations, which we developed for modern computers. \cite{2021simi.conf...86R} laid down another modern solution of the problem. The general results of Bo\v{s}kovi\'{c}’s determinations of sunspot positions, and then solar rotation elements $\Omega$, $i$, $T'$, and $T''$ were summarized in \cite{2023SoPh..298..122H}. In Table 7 of that paper results presented by \cite{1978A&A....62..165W} were exteded with the Bo\v{s}kovi\'{c}’s determinations and with the results published after 1978.

\cite{1785BoscovichII} described solution for solar rotation elements using his methodology of arithmetic means, as well as planar geometrical construction, trigonometric planar solution, and trigonometric spherical solution. 
The methodology of arithmetic means calculates the solar rotation elements $\Omega$, $i$, $T'$, and $T''$ separately. The trigonometric planar solution calculates $\Omega$ and $i$ together using three sunspot positions. The trigonometric planar solution calculates $T'$ and $T''$ in the same way as the methodology of arithmetic means. 

The last one, the trigonometric spherical solution, was only described, but Bo\v{s}kovi\'{c} did not develop the equations for the solution of the method. Bo\v{s}kovi\'{c} named this solution \textit{very long and unpractical beside his simpler graphical solution (geometric construction) and trigonometric planar solution} \citep[\textnumero81]{1785BoscovichII}. Bo\v{s}kovi\'{c}'s argument is valid for for trigonometric and logarithmic calculation he used then.%, which we do not use any more.

\begin{sidewaysfigure}
    \centering
    \includegraphics[width=0.3\textheight%, angle=90
    ]{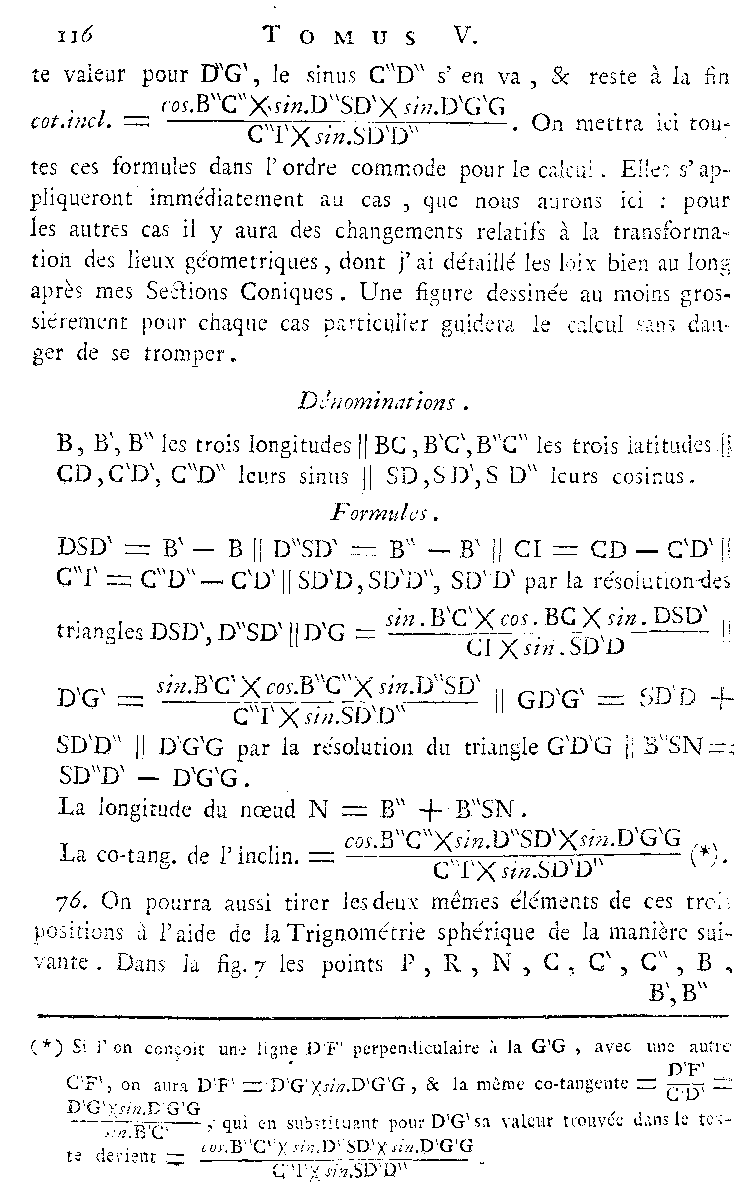}
    \includegraphics[width=0.3\textheight%, angle=90
    ]{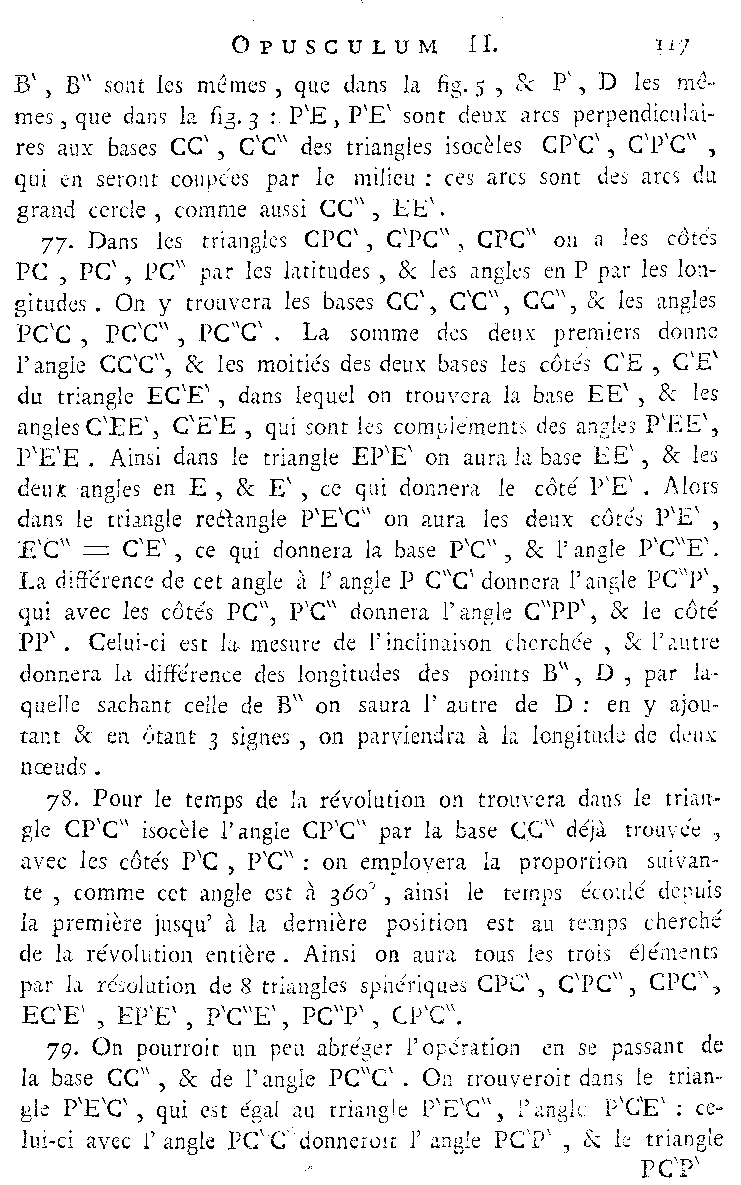}
    \includegraphics[width=0.3\textheight%, angle=90
    ]{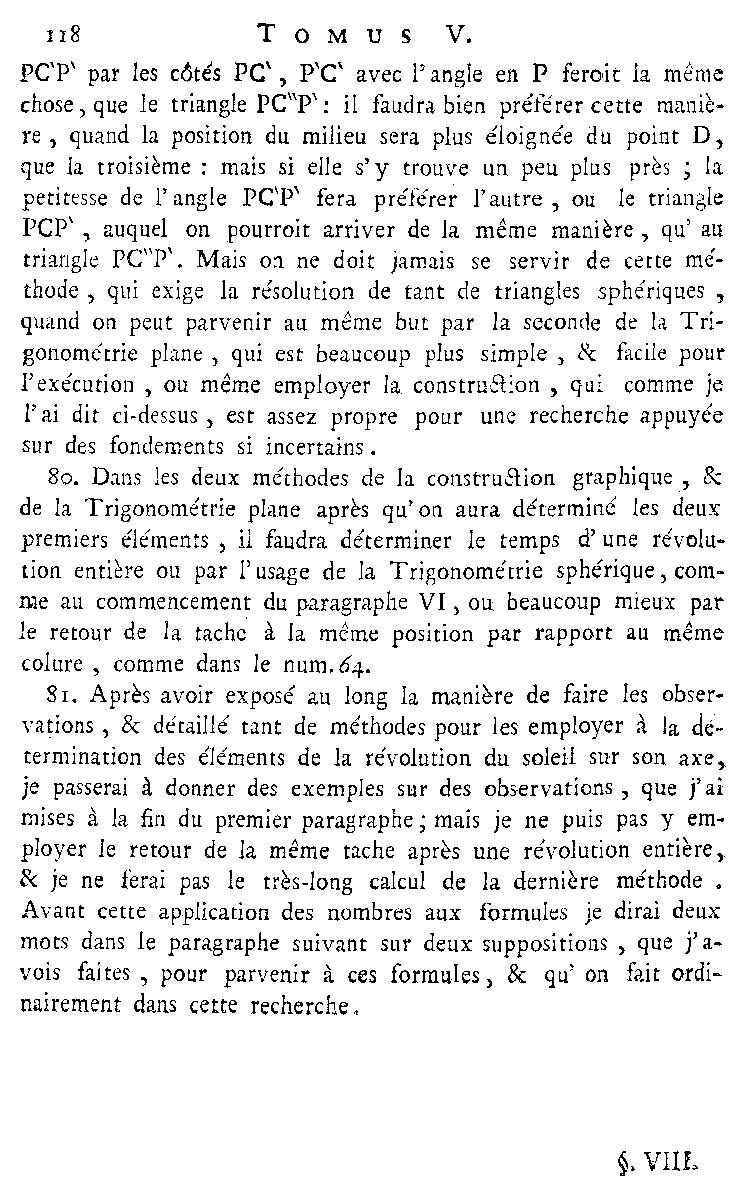}
    \caption{Description of the trigonometric spherical solution \citep[\S VII., \textnumero76-81, p116-118]{1785BoscovichII}{}.}% MAYBE PUT IN THE APPENDIX OF THE PAPER?
    \label{fig:p116_117_118}
\end{sidewaysfigure}

In the present work, we followed Bo\v{s}kovi\'{c}'s descriptions in \S.VII., \textnumero76-\textnumero 81 (Figure~\ref{fig:p116_117_118}) to develop the equations for the trigonometric spherical solution \cite[\S.VII., \textnumero76-\textnumero 81]{1785BoscovichII}. Moreover, we adapted the equations for modern computers. In the present work, we recalculated Bo\v{s}kovi\'{c}'s original example with here developed and adapted equations. The importance of the trigonometric spherical solution is that it calculates all three solar rotation elements $\Omega$, $i$, and $T'$ with three positions of the same sunspot in a single procedure.
\section{Trigonometric spherical solution for $i$, $\Omega$, and $T'$ (Methods)}
\label{s:Methods}
\begin{figure}[h]
    \centering
    \includegraphics[width=0.6\textwidth, angle=90]{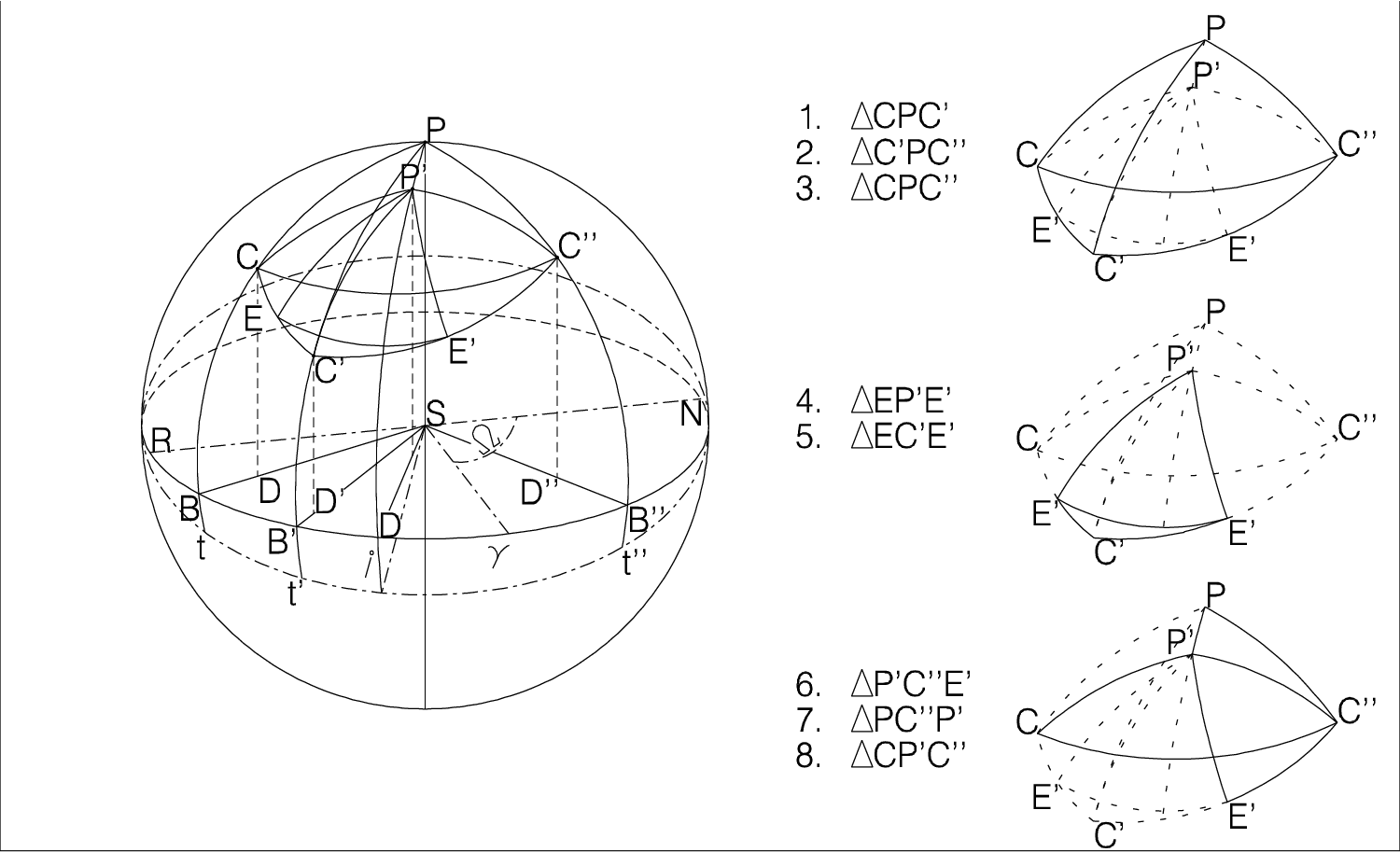}%8triangles.eps}
    \caption{Three sunspot positions $(B,C,t)$, $(B',C',t')$, and $(B'',C'',t'')$ in ecliptic coordinates $B,C$ and mean solar time $t$ with poles $P$ and $P'$ make eight triangles for the trigonometric spherical solution by Bo\v{s}kovi\'{c} for solar rotation elements: the inclination of the solar equator regarding ecliptic $i$, the longitude of the ascending node $\Omega$, and the sidereal solar rotation rate $T'$ \citep{1785BoscovichII}.}
    \label{fig:8triangles}
\end{figure}

Trigonometric spherical solution for $i$, $\Omega$, and $T'$ is the third solution besides two solutions for $\Omega$ and $i$: the graphical solution (geometric construction) and the trigonometric planar solution \citep[\S VII., \textnumero67-\textnumero75 and \S XIII., \textnumero 129-\textnumero140]{1785BoscovichII}. The third solution of the method, developed here with equations, gives us all three Carrington's solar rotation elements using three sunspot positions and its mean solar time: %solar equator inclination
$i$, %longitude of the ascending node 
$\Omega$, and %sidereal solar rotation rate 
$T'$.
\cite{1785BoscovichII} described the solution in \S VII, \textnumero 76 - \textnumero 78  using eight spherical triangles defined with the northern ecliptic pole $P$, the northern equator's pole $P'$, three sunspot positions $C(B,C)$, $C'(B',C')$, and $C''(B'',C'')$ in ecliptic coordinate system are presented in Figure~\ref{fig:8triangles}, where $B$ denotes ecliptic longitude and $C$ ecliptic latitude.

There are three groups of triangles (Figure~\ref{fig:8triangles}):
\begin{enumerate}
    \item Sunspots $C$, $C'$, and $C''$ with ecliptic pole $P$ make three triangles $\triangle_1 CPC'$,  $\triangle_2 C'PC''$, and $\triangle_3 CPC''$.
    \item Midpoints of $CC'$ and $C'C''$, $E$ and $E'$ respectively with equator's pole $P'$ make two triangles: $\triangle_4 EP'E'$ and $\triangle_5 EC'E'$.
    \item Side $P'C''$ with sunspot $C$, ecliptic pole $P$, and $E'$ as midpoint of $C'C''$ make three triangles: $\triangle_6 P'C''E'$, $\triangle_7 P'C''P$, and $\triangle_8 CP'C''$.
\end{enumerate}

The solution of mentioned eight oblique spherical triangles gives $i$, $\Omega$, and $T'$ using the only one solution of the method. These triangles could be solved using so-called \textit{unfulfilled solution} (it uses whole triangle's angles and sides) or \textit{fulfilled solution} (it uses half-sums and half-differences of triangle's angles and sides) of spherical triangles. We used the former solution in the present work as follows.

\subsection{Solution for $i$}
\label{ss:i}
In the triangle $\triangle_1 CPC'$, we determine the side $CC'$ using the cosine rule for side $CC'$ with the known opposite angle $(B'-B)$ and its sides $(90^{\circ}-C)$ and $(90^{\circ}-C')$ as follows
$$\cos CC' =  \cos (90^{\circ}-C) \cdot \cos (90^{\circ}-C')+\sin (90^{\circ}-C) \cdot \sin (90^{\circ}-C') \cdot \cos (B'-B)$$
\begin{equation}
    %\alpha_1
    \label{eq:c0c1}
    \cos CC' = \sin C \cdot \sin C' + \cos C \cdot \cos C' \cdot \cos (B'-B).
\end{equation}
Similarly, for the triangles %$\triangle_2$ and $\triangle_3$, 
$\triangle_2 C'PC''$ and $\triangle_3 CPC''$ we have $C'C''$ and $C''C$, respectively
\begin{equation}
    %\alpha_2
    \label{eq:c1c2}
    \cos C'C'' = \sin C' \cdot \sin C'' + \cos C' \cdot \cos C'' \cdot \cos (B''-B')
\end{equation}
\begin{equation}
    %\alpha_3
    \label{eq:c2c0}
    \cos C''C = \sin C'' \cdot \sin C + \cos C'' \cdot \cos C \cdot \cos (B''-B).
\end{equation}
In the triangles $\triangle_1$, $\triangle_2$, and $\triangle_3$ we know all the three sides and now we can find other angles\footnote{In equations we assign angles as follows, for example $\angle PC'C$ has vertex in $C'$ with sides $PC'$ and $C'C$, as $PC'C$, without sign $\angle$.} using the cosine rule as follows
$$\cos PC'C = \frac{\cos (90^{\circ}-C) - \cos (90^{\circ}-C') \cdot \cos CC'}{\sin (90^{\circ}-C') \cdot \sin CC'}$$
\begin{equation}
    %\beta_1
    \label{eq:p0c1c0} %4
    \cos PC'C = \frac{\sin C - \sin C' \cdot \cos CC' }{\cos C' \cdot \sin CC'}.
\end{equation}
Similarly, we have in triangles $\triangle_2$ and $\triangle_3$, angles $PC'C''$ and $PC''C'$, respectively
%%1%$$\cos PC'C'' = \frac{\cos (90^{\circ}-C'') - \cos (90^{\circ}-C') \cdot \cos C'C''}{\sin (90^{\circ}-C') \cdot \sin C'C''}$$
\begin{equation}
    %\beta_2 5
    \label{eq:p0c1c2}
    \cos PC'C'' = \frac{\sin C'' - \sin C' \cdot \cos C'C'' }{\cos C' \cdot \sin C'C''}
\end{equation}
%%2%$$\cos PC''C' = \frac{\cos (90^{\circ}-C') - \cos (90^{\circ}-C'') \cdot \cos C'C''}{\sin (90^{\circ}-C'') \cdot \sin C'C''}$$
\begin{equation}
    %\beta_3
    \label{eq:p0c2c1}
    \cos PC''C' = \frac{\sin C' - \sin C'' \cdot \cos C'C'' }{\cos C'' \cdot \sin C'C''}.
\end{equation}
In the triangle $\triangle CC'C''$ the angle by $C'$ is the sum of the angles in Equations~\ref{eq:p0c1c0} and \ref{eq:p0c1c2}
\begin{equation}
    \label{eq:c0c1c2}
    \angle CC'C'' = \angle PC'C + \angle PC'C''.
\end{equation}

The triangle of three sunspot positions $\triangle CC'C''$ has all three known sides (Equations~\ref{eq:c0c1}, \ref{eq:c1c2}, and \ref{eq:c2c0}). From the triangle $\triangle CC'C''$ we can get all its angles using the cosine rule, because we have all cosines of all sides in Equations~\ref{eq:c0c1}, \ref{eq:c1c2}, and \ref{eq:c2c0}
\begin{equation}
    %\gamma_1
    \label{eq:c2c0c1}
    \cos C''CC' = \frac{\cos C'C'' - \cos C''C \cdot \cos CC'}{\sin C''C \cdot \sin CC'}
\end{equation}
\begin{equation}
    %\gamma_2
    \label{eq:c0c1c2_cos}
    \cos CC'C'' = \frac{\cos CC'' - \cos CC' \cdot \cos C'C''}{\sin CC' \cdot \sin C'C''}
\end{equation}
\begin{equation}
    %\gamma_3
    \label{eq:c1c0c2}
    \cos C'C''C = \frac{\cos CC' - \cos CC'' \cdot \cos C'C'' }{\sin C''C \cdot \sin C'C''}.
\end{equation}
Ru\dj er Bo\v{s}kovi\'{c} used the sum of angles in Equations~\ref{eq:p0c1c0} and \ref{eq:p0c1c2} which we determined in Equation~\ref{eq:c0c1c2}.

%Angles i
In the triangle $\triangle_3 CPC''$ we have %could solve using%:
%\begin{enumerate}
%    \item C
three known sides: $90^\circ-C$, $90^\circ-C''$, and $CC''$ (Equation~\ref{eq:c2c0}).
%    \item Tangent rule of half angle.
%\end{enumerate}
For determination of the angles $\angle PCC''$ and $\angle PC''C$ we use the cosine rule% which we used before.%. We do not need this angles for this research, so we skip this determination.
$$\cos PCC'' = \frac{\cos (90^{\circ}-C'') - \cos (90^{\circ}-C) \cdot \cos CC''}{\sin (90^{\circ}-C) \cdot \sin CC''}$$
\begin{equation}
    %\beta_
    \label{eq:p0c0c2}
    \cos PCC'' = \frac{\sin C'' - \sin C \cdot \cos CC'' }{\cos C \cdot \sin CC''}.
\end{equation}
%where $\cos CC''$ is determined with Equation~\ref{eq:c2c0}. 
In similar way, we determine the angle in $C''$
$$\cos PC''C = \frac{\cos (90^{\circ}-C) - \cos CC'' \cdot \cos (90^{\circ}-C'')}{\sin CC'' \cdot \sin (90^{\circ}-C'')}$$
\begin{equation}
    %\beta_
    \label{eq:p0c2c0}
    \cos PC''C = \frac{\sin C - \sin C'' \cdot \cos CC'' }{\cos C'' \cdot \sin CC''}.
\end{equation}
The angles $\angle PC'C$ and $\angle PC''C$ can be found using the cotangent rule for $\angle PC'C$ from $\triangle_1 CPC'$
$$\cot PCC' \cdot \sin (B'-B) = \cot (90^\circ - C') \sin (90^\circ - C) - \cos (B'-B) \cdot \cos (90^\circ - C)$$
$$\frac{\sin (B'-B)}{\tan PCC'} = \tan C' \cdot \cos C - \cos (B'-B) \cdot \sin C $$
\begin{equation}
    %\delta_1\
    %\label{eq:delta_1}
    \label{eq:p0c0c1_tan}
    \tan PCC' = \sin (B'-B) \cdot [\tan C' \cdot \cos C - \cos (B'-B) \cdot \sin C ]^{-1}.
\end{equation}
Similarly, angles $PC'C$ and $PC'C''$ are:
$$\cot PC'C \cdot \sin PCC' = \cot (90^\circ - C) \sin CC' - \cos PCC' \cdot \cos CC')$$
%%3%$$\frac{\sin PCC'}{\tan PC'C} = \tan C \cdot \sin CC' - \cos PCC' \cdot \cos CC'$$
\begin{equation}
    %\delta_2\
    %\label{eq:delta_2}
    \label{eq:p0c1c0_tan}
    \tan PC'C = \sin PCC' \cdot [\tan C \cdot \sin CC' - \cos PCC' \cdot \cos CC']^{-1}
\end{equation}
$$\cot PC'C'' \cdot \sin (B''-B') = \cot (90^\circ - C'') \sin (90^\circ - C') - \cos (B''-B') \cdot \cos (90^\circ - C')$$
%%4%$$\frac{\sin (B''-B')}{\tan PC'C''} = \tan C'' \cdot \cos C' - \cos (B''-B') \cdot \sin C' $$
\begin{equation}
    %\delta_3\
    %\label{eq:delta_3}
    \label{eq:p0c1c2_tan}
    \tan PC'C'' = \sin (B''-B') \cdot [\tan C'' \cdot \cos C' - \cos (B''-B') \cdot \sin C']^{-1}.
\end{equation}
The cotangent rule applied in Equations~\ref{eq:p0c0c1_tan}, \ref{eq:p0c1c0_tan}, and \ref{eq:p0c1c2_tan} for calculation use only sunspot coordinates - source input values, so they are better for computer calculation then the cosine rule Equations~\ref{eq:p0c1c0},~\ref{eq:p0c1c2}, and~\ref{eq:p0c2c1}.

The second group of triangles $\triangle_4$ and $\triangle_5$ presented in Figure~\ref{fig:8triangles} are $\triangle_4 EP'E'$ and $\triangle_5 EC'E'$, where $E$ is the midpoint of the side $CC'$ and $E'$ is the midpoint of the side $C'C''$. We have 
\begin{equation}
\label{eq:e0c1}
CE=EC'=\frac{CC'}{2}
\end{equation}
and
\begin{equation}
\label{eq:e1c2}
C'E'=E'C''=\frac{C'C''}{2}.
\end{equation}
The midpoints $E$ and $E'$ make two triangles, the first one with the equator's pole $P'$ and the second one with the middle sunspot position $C'$.

Bo\v{s}kovi\'{c}'s description of this solution in \textnumero 76 and \textnumero 77 separates triangle $\triangle CP'C'$ in two right-angle triangles $\triangle CP'E$ and $\triangle EP'C'$, where the side $EP'$ is perpendicular to the side $CC'$. 

The base side $EE'$ of the $\triangle_5 EC'E'$ we determine with the angle $\angle CC'C''$ and the sides $EC'$ and $C'E'$ using the cosine rule
\begin{equation}
\label{eq:e0e1}
    \cos EE' = \cos EC' \cdot \cos C'E' + \sin EC' \cdot \sin C'E' \cdot \cos CC'C''.
\end{equation}

Equation~\ref{eq:e0e1} uses $\angle CC'C''$, which Bo\v{s}kovi\'{c} determined in Equation~\ref{eq:c0c1c2}, but it can be solved using Equation~\ref{eq:c0c1c2_cos}, too. In the triangle $\triangle_5 EC'E'$ we determined all three sides, so we can determine its angles in $E$ and $E'$ using the cosine rule
\begin{equation}
    \label{eq:c1e0e1}
    \cos C'EE' = \frac{\cos C'E' - \cos EE' \cdot \cos EC'}{\sin EE' \cdot \sin EC'}
\end{equation}
\begin{equation}
    \label{eq:c1e1e0}
    \cos C'E'E = \frac{\cos C'E- \cos EE' \cdot \cos E'C'}{\sin EE' \cdot \sin E'C'}.
\end{equation}

The angles $\angle C'EE'$ and $\angle C'E'E$ are complements of the angles $\angle P'EE'$ and $\angle P'E'E$, respectively
\begin{equation}
\label{eq:p1e0e1}
    \angle P'EE' = 90^{\circ} - \angle C'EE'
\end{equation}
\begin{equation}
\label{eq:p1e1e0}
    \angle P'E'E = 90^{\circ} - \angle C'E'E.
\end{equation}
The solution for $\angle EP'E'$ using the polar cosine rule is
$$\cos EP'E' = - \cos P'EE' \cdot \cos EE'P' + \sin P'EE' \cdot \sin EE'P' \cdot \cos EE' $$
$$\cos EP'E' = - \cos (90^\circ - C'EE') \cdot \cos (90^\circ - C'E'E) +$$
$$+ \sin (90^\circ - C'EE') \cdot \sin (90^\circ - C'E'E) \cdot \cos EE' $$
\begin{equation}
    \label{eq:e0p1e1}
    \cos EP'E' = - \sin C'EE' \cdot \sin C'E'E + \cos C'EE' \cdot \cos C'E'E \cdot \cos EE'.
\end{equation}
From the triangle $\triangle_4 EP'E'$ and Equations \ref{eq:e0e1}, \ref{eq:p1e0e1}, and \ref{eq:e0p1e1} by using the sine rule we get
$$\frac{\sin P'EE'}{\sin P'E'} = \frac{\sin EP'E'}{\sin EE'}$$
$$\sin P'E' = \sin EE' \cdot \frac{\sin P'EE'}{\sin EP'E'} = \sin EE' \cdot \frac{\sin (90^{\circ} - C'EE')}{\sin EP'E'}$$
\begin{equation}
    \label{eq:p1e1}
    \sin P'E' = \frac{\sin EE' \cdot \cos C'EE'}{\sin EP'E'}.
\end{equation}

In the triangle $\triangle_4 EP'E'$ we determined the side $EE'$ and the angles on it which are the complements of the angles of the triangle $\triangle_5 EC'E'$. From this we can determine the side $P'E'$ using the cotangent rule:

$$\cot P'E' \cdot \sin EE' = \cot (90^\circ - C'EE') \cdot \sin (90^\circ - C'E'E) +$$
$$+ \cos (90^\circ - C'E'E) \cdot \cos EE' $$
%%5%$$\frac{\sin EE'}{\tan P'E'} = \tan C'EE' \cdot \cos C'E'E + \sin C'E'E \cdot \cos EE'$$
\begin{equation}
    \label{eq:p1e1_tan}
    \tan P'E' = \sin EE' \cdot [\tan C'EE' \cdot \cos C'E'E + \sin C'E'E \cdot \cos EE']^{-1}.
\end{equation}

In the right-angle triangle $\triangle_6 P'E'C''$ we know two sides $E'C''=E'C'$ and $P'E'$ (Equations~\ref{eq:e1c2} and ~\ref{eq:p1e1_tan}), so we can determine the side $P'C''$ using the cosine rule and the angle $\angle P'C''E'$ for the right-angle triangle
\begin{equation}
    \label{eq:p1c2}
    \cos P'C'' = \cos P'E' \cdot \cos E'C''
\end{equation}
\begin{equation}
    \label{eq:p1c2e1}
    \tan P'C''E' = \frac{\tan P'E'}{\sin E'C''}
\end{equation}
\begin{equation}
   \label{eq:p0c2p1}
    \angle PC''P' = \angle PC''C' - \angle P'C''E',
\end{equation}
where we determined $PC''C' = PC''E'$ in Equation~\ref{eq:p0c2c1}.

In the triangle $\triangle_7 PC''P'$ we know two sides $PC''$ and $P'C''$ and the angle between them $\angle PC''P'$ (Equation~\ref{eq:p0c2p1}), so we can determine the side $PP'=i$ by the cosine rule
$$\cos PP' = cos P'C'' \cdot \cos (90^{\circ}-C'') + \sin P'C'' \cdot \sin (90^{\circ}-C'') \cdot \cos PC''P'$$
\begin{equation}
  \label{eq:p0p1_i}
    \cos PP' = cos P'C'' \cdot \sin C'' + \sin P'C'' \cdot \cos C'' \cdot \cos PC''P',
\end{equation}
where $PP'=i$ is solar equator inclination.

\subsection{Solution for $\Omega$}
\label{ss:Omega}
We use the same triangle $\triangle_7 PC''P'$ for determination of the angle $\angle (B''-D) = \angle P'PC''$ in the northern ecliptic pole $P$ using the cosine rule and Equations~\ref{eq:p1c2}, \ref{eq:p0c2p1}, and~\ref{eq:p0p1_i}
$$\cos (B''-D) = \frac{\cos P'C'' - \cos i \cdot \cos (90^{\circ}-C'')}{\sin i \cdot \sin (90^{\circ}-C'')}$$
\begin{equation}
    \label{eq:b2d0}
    \cos (B''-D) = \frac{\cos P'C'' - \cos i \cdot \sin C''}{\sin i \cdot \cos C''}
\end{equation}
\begin{equation}
    \label{eq:d}
    D=B''-(B''-D).
\end{equation}
As \cite{1785BoscovichII} described in \textnumero 77, the longitude of the ascending node $N$ and the longitude of the descending node $R$ we determine by adding and subtracting three Zodiac signs ($1^s=30^\circ$, $3^s=90^\circ$) to $D$
\begin{equation}
    \label{eq:r}
    R=D+3^s=D+90^\circ=[B''-(B''-D)]+90^\circ
\end{equation}
\begin{equation}
    \label{eq:n_Omega}
    N=D-3^s=D-90^\circ=[B''-(B''-D)]-90^\circ.
\end{equation}
%where $N$ is the longitude of ascending node and $R$ is the longitude of descending node. 
Ru\dj er Bo\v{s}kovi\'{c} denoted longitude of ascending node with $N$, today we denote it with $\Omega$.
\subsection{Solution for $T'$}
\label{ss:T}
In \textnumero 78 \cite{1785BoscovichII} determined sidereal solar rotation rate from the isosceles triangle $\triangle_8 CP'C''$. The angle $\angle CP'C''$ we determine using three sides $CC''$ and $P'C=P'C''$ (Equation~\ref{eq:c2c0} and~\ref{eq:p1c2}) by the cosine rule
$$\cos CP'C'' = \frac{\cos CC'' - \cos CP' \cdot \cos P'C''}{\sin CP' \cdot \sin P'C''}$$
\begin{equation}
    \label{eq:c0p1c2}
    \cos CP'C'' = \frac{\cos CC'' - (\cos P'C'')^2}{(\sin P'C'')^2}.
\end{equation}
\cite{1785BoscovichII} put the ratio
\begin{equation}
    \label{eq:tratio}
    \angle CP'C'' : 360^\circ = \Delta t : T'
\end{equation}
\begin{equation}
    \label{eq:t1_sid}
    T' = \Delta t \cdot \frac{360^\circ}{\angle CP'C''},
\end{equation}
where $T'$ is the sidereal solar rotation rate and $\Delta t_{13} = t'' - t$ is the difference of mean solar times of the third and the first sunspot position, $t$ is the mean solar time of the first sunspot position, and $t''$ of the third sunspot position.

In the footnote of \textnumero 66 \cite{1785BoscovichII} determined %equation for 
synodic period
\begin{equation}
    \label{eq:t2_syn}
    T''=\frac{A \cdot T'}{A - T'},
\end{equation}
where $A=365.25$ days.

The calculation of the solar rotation parameters using Bo\v{s}kovi\'{c}'s sunspot positions (Table~\ref{tbl:sunspots}) %measurements 
with the described trigonometric spherical solution method is presented in Table~\ref{tbl:calculation}.
\subsection{Trigonometric spherical short solution}
\label{ss:short}
Ru\dj er Bo\v{s}kovi\'{c} performed the Trigonometric spherical short solution (TSSS) of the method \citep[\textnumero 79]{1785BoscovichII}{}. The trigonometric spherical short solution uses spherical triangles $\triangle_2 C'PC''$, $\triangle_6 P'E'C''$, $\triangle_7 PC''P'$, $\triangle_9 P'E'C'$, and $\triangle_{10} PC'P'$ (Figure~\ref{fig:8triangles}). 
\subsubsection{The short solution equation development}
\label{sss:eq_dev}
The short solution begins with the side $C'C''$\footnote{In the \textnumero79 instead the side $CC''$ should be the side $C'C''$.} and the angle $\angle PC''C'$ in the $\triangle_2 C'PC''$. In $C'$ we have the angle $\angle PC'C''$ (Equation~\ref{eq:p0c1c2_tan}). In  $\triangle_2 C'PC''$ we are looking for the angle $\angle PC''C'$ by $C''$, which we can find using the cotangent rule
$$\cot PC''C' \cdot \sin  PC'C'' = \cot (90^\circ - C') \cdot \sin C'C'' - \cos  PC'C'' \cdot \cos C'C''$$
%%6%$$\frac{\sin  PC'C''}{\tan PC''C'} = \tan C' \cdot \sin C'C'' - \cos  PC'C'' \cdot \cos C'C''$$
\begin{equation}
    \label{eq:p0c2c1_tan}
    \tan PC''C' = \sin  PC'C'' \cdot [\tan C' \cdot \sin C'C'' - \cos  PC'C'' \cdot \cos C'C'']^{-1},
\end{equation}
where $C'$ is the ecliptic latitude of the middle sunspot position and we know the side $C'C''$ and the angle $\angle PC'C''=\angle P'C'E'$ (Equations~\ref{eq:c1c2} and~\ref{eq:p0c1c2_tan}). The midpoint of the side $C'C''$ is $E'$ (Equations~\ref{eq:c1c2} and~\ref{eq:e1c2}). 

The triangles $\triangle_9 P'E'C'$ and $\triangle_6 P'E'C''$ are mirroring (symmetric) regarding the side $P'E'$ with the right angle in $E'$, so $P'C'=P'C''$. The side $P'E'$ of the triangle $\triangle_6 P'C''E'$ and $\triangle_9 P'C'E'$ we determine with:
\begin{enumerate}
    \item the cotangent rule with Equation~\ref{eq:p1e1_tan}, which uses Equations~\ref{eq:e0e1},~\ref{eq:c1e0e1}, and~\ref{eq:c1e1e0}, and
    \item the sine rule with Equation~\ref{eq:p1e1}, which uses Equations~\ref{eq:e0e1},~\ref{eq:c1e0e1}, and~\ref{eq:e0p1e1} (Equation~\ref{eq:e0p1e1} uses~\ref{eq:e0e1},~\ref{eq:c1e0e1}, and~\ref{eq:c1e1e0}).
\end{enumerate}
The cotangent rule solution for $P'E'$ is a little simpler than the sine rule solution.

We are looking for $P'C''$ in the right-angle triangle $\triangle_6 P'E'C''$. The side $P'C''=P'C'$ of the $\triangle_6$ we solved with Equation~\ref{eq:p1c2}.

In the triangles $\triangle_9 P'E'C' = \triangle_6 P'E'C''$ we look for the angles in $C'$ and $C''$. The angle $\angle P'C''E'$ in $C''$ we solve with Equation~\ref{eq:p1c2e1}. The angle $\angle P'C'E'$ in $C'$ we can solve in the same way
\begin{equation}
    \label{eq:p1c1e1}
    \tan P'C'E' = \frac{\tan P'E'}{\sin E'C'}.
\end{equation}
The triangle $\triangle_{10} PC'P'$ we can solve using the sides $PC'=90^\circ - C'$ and $P'C'$ with the angle $\angle PC'P'$ between them in $C'$
\begin{equation}
    \label{eq:p0c1p1}
    \angle PC'P' = \angle PC'C'' - \angle P'C'C''.
\end{equation}
The solution of $\triangle_{10} PC'P'$ gives the third side $PP'$ using the cosine rule for the sides 
$$\cos PP'= \cos (90^\circ - C') \cdot \cos P'C' + \sin (90^\circ - C') \cdot \sin P'C' \cdot cos PC'P'$$
\begin{equation}
    \label{eq:p0p1_shortc1}
    \cos PP'= \sin C' \cdot \cos P'C' + \cos C' \cdot \sin P'C' \cdot \cos PC'P',
\end{equation}
where $PP'=i_{Short_{C'}}$ is the solar equator inclination calculated in $\triangle_{10} PC'P'$.

The solution of $\triangle_{10} PC'P'$ gives the angle $\angle C'PP'=\angle (D-B')$ using the cosine rule for the sides
$$\cos P'C' = \cos (90^\circ - C') \cdot \cos PP' + \sin (90^\circ - C') \cdot \sin PP' \cdot \cos (D-B')$$
%%7%$\cos P'C' - \sin C' \cdot \cos PP' = \cos C' \cdot \sin PP' \cdot \cos (D-B')$$
\begin{equation}
    \label{eq:d0b1}
    \cos (D-B')=\frac{\cos P'C' - \sin C' \cdot \cos PP'}{\cos C' \cdot \sin PP'}.
\end{equation}
We have the longitude of the maximal latitude of the sunspot over ecliptic $D$
$$D=(D-B')+B',$$
and then we can calculate the longitude of the ascending node $N=\Omega=D-\textit{}90^\circ$ and the longitude of the descending node $R=D+90^\circ$ (Equations~\ref{eq:n_Omega} and~\ref{eq:r}).

Sidereal rotational rate $T'$ we calculate like in Subsection~\ref{ss:T} from $\triangle_8 CP'C''$ using the cosine rule

$$\cos C'P'C'' = \frac{\cos C'C'' - \cos C'P' \cdot \cos P'C''}{\sin C'P' \cdot \sin P'C''},$$
where $C'P' = P'C''$, and
\begin{equation}
    \label{eq:c1p1c2}
    \cos C'P'C'' = \frac{\cos C'C'' - (\cos P'C')^2}{(\sin P'C')^2}.
\end{equation}
Using Equation~\ref{eq:tratio} we have $T'$ using Equation~\ref{eq:t1_sid} where $\Delta t_{23}=t''-t'$.

The calculation of described \textit{Trigonometric spherical short solution} is presented in Table~\ref{tbl:calculation_short}.

The second solution for $PP'=i_{Short_{C''}}$ is in $C''$ from the triangle $\triangle_7 PC''P'$. The triangles $\triangle_9 P'E'C'$ and $\triangle_6 P'E'C''$ are mirroring regarding $P'E'$, so angles $\angle P'C''E' = \angle P'C'E'$ (Equation~\ref{eq:p1c1e1}) and the sides $P'C'' = P'C'$ (Equation~\ref{eq:p1c2}) are equal. We calculate the angle $\angle PC''P'$ in $C''$
\begin{equation}
    \label{eq:p0c2p1_short}
    \angle PC''P' = \angle PC''E' - \angle P'C''E',
\end{equation}
where $\angle PC''E'=\angle PC''C'$ (Equation~\ref{eq:p0c2c1_tan}).
%and $P'C''E' = P'C'E'$ (Equation~\ref{eq:p1c1e1}) because triangles $\triangle_9 P'E'C'$ and $\triangle_6 P'E'C''$ are mirroring regarding $P'E'$.

In the triangle $\triangle_7 PC''P'$ we have the sides $PC''=90^\circ - C''$ and $P'C'' = P'C'$ (Equation~\ref{eq:p1c2}) 
%look for side $P'C''$
from neighboring triangle $\triangle_6 P'E'C''$. The $\angle PC''P'$ between the sides in $C''$ is already determined in Equation~\ref{eq:p0c2p1} and here we use the signs for the short solution (Equation~\ref{eq:p0c2p1_short}). The second result of the short solution we determine in Equation~\ref{eq:p0p1_i}.
\subsubsection{The short solution equation development - another solution}
\label{sss:eq_dev_an}
\cite{1785BoscovichII} suggested in \textnumero 79 another solution, which starts with the triangle $\triangle_{11} PCP'$. He discussed the longitudes of sunspots $B$, $B'$, and $B''$: he described the procedure when the position of means is more distant from $D$ then third sunspot position,
\begin{equation}
    \label{eq:deltab_more}
    |\Delta \bar{B}| > |\Delta B''|,
\end{equation}
where $\bar{B}=(B+B')/2$, $\Delta \bar{B} = \bar{B} - D$, and $\Delta B'' = B'' - D$, but if it is less distant
\begin{equation}
    \label{eq:deltab_less}
    |\Delta \bar{B}| < |\Delta B''|,
\end{equation}
then the angle $\angle PC'P'$ will be so small that we will prefer the solution starting with the triangle $\triangle_{11} PCP'$, which we develop like the short solution already described.

Another solution of Trigonometric spherical short solution equations development is similar to the just described one, it uses the triangles $\triangle_{11} PCP'$, $\triangle_{12} P'EC$, and $\triangle_{13} P'EC'$. We should find $PP'=i$ from the triangle $\triangle_{11} PCP'$ using the cosine rule, so we will need the angle $\angle PCP'$ and the sides $PC=90^\circ-C$, as well as the side $P'C$. The angle $\angle PCP'$ is then
\begin{equation}
    \label{eq:p0c0p1}
    \angle PCP' = \angle PCC' - \angle P'CC',
\end{equation}
where $\angle PCC' = \angle PCE$ ($E$ is midpoint of the side $CC'$).

We are looking for $\angle PCC'$ and the side $P'C$ from two triangles with right angle in $E$, $\triangle_{12} P'EC$ and $\triangle_{13} P'EC'$. They are mirroring regarding the side $P'E$ which is perpendicular to the side $CC'$. We can determine the sides $P'C=P'C'$ from mirroring right angle triangles $\triangle_{12} P'EC$ and $\triangle_{13} P'EC'$. We are looking for $P'E$ using the sine rule and Equation~\ref{eq:p1e1e0}

$$\frac{\sin EP'E'}{\sin EE'}=\frac{\sin P'E'E}{\sin P'E}$$
$$\sin P'E=\sin EE' \cdot \frac{\sin P'E'E}{\sin EP'E'}=\sin EE' \cdot \frac{\sin (90^\circ - C'E'E)}{\sin EP'E'}$$
\begin{equation}
    \label{eq:p1e0}
    \sin P'E=\sin EE' \cdot \frac{\cos C'E'E}{\sin EP'E'}.
\end{equation}
The side $P'E$ we can determine from the $\triangle_4 EP'E'$ using the cotangent rule, as we did for the side $P'E'$ (Equation~\ref{eq:p1e1_tan})
$$\cot (90^\circ - CE'E) \cdot \sin (90^\circ - CEE') = \cot P'E \cdot \sin EE' - \cos (90^\circ - CEE') \cdot \cos EE'$$
$$\tan CE'E \cdot \cos CEE' +  \sin CEE' \cdot \cos EE'= \cot P'E \cdot \sin EE'$$
\begin{equation}
    \label{eq:p1e0_tan}
    \tan P'E = \sin EE' \cdot [\tan CE'E \cdot \cos CEE' +  \sin CEE' \cdot \cos EE']^{-1}.
\end{equation}
In the right angle triangles $\triangle_{12} P'EC$ and $\triangle_{13} P'EC'$ we can determine sides $P'C$ and $P'C'$. We know sides $P'E$ and $CE=EC'$ (Equation~\ref{eq:e0c1}), so we use the cosine rule for the right angle triangles $\triangle_{12}$ and $\triangle_{13}$
\begin{equation}
    \label{eq:p1c0}
    \cos P'C = \cos P'E \cdot \cos EC
\end{equation}
\begin{equation}
    \label{eq:p1c1}
    \cos P'C' = \cos P'E \cdot \cos EC'.
\end{equation}
These sides should be equal to $P'C''$ (Equation~\ref{eq:p1c2}). Equation development for $P'C$ and $P'C'$ is similar as before (Equations~\ref{eq:e0c1} to~\ref{eq:p1c2}).

The angle $\angle P'CC'=\angle P'C'C$ we can get from right angle triangle $\triangle_{12} P'EC$ or $\triangle_{13} P'EC'$
\begin{equation}
    \label{eq:p1c0e0}
    \tan P'CE=\frac{\tan P'E}{\sin EC}
\end{equation}
\begin{equation}
    \label{eq:p1c1e0}
    \tan P'C'E=\frac{\tan P'E}{\sin EC'},
\end{equation}
where $\angle P'CC'= \angle P'CE$ and $\angle P'C'C= \angle P'C'E$ ($E$ is the midpoint of $CC'$).

The solar equator inclination from $\triangle_{11} PCP'$ is
$$\cos PP' = \cos (90^\circ - C) \cdot cos P'C +\sin  (90^\circ - C) \cdot \sin P'C \cdot \cos PCP'$$
\begin{equation}
    \label{eq:p0p1_shortc0}
    \cos PP' = \sin C \cdot cos P'C + \cos C \cdot \sin P'C \cdot \cos PCP',
\end{equation}
where $PP'=i_{Short_{C}}$ is solar equator inclination.

The longitude of the ascending node we can find from the same triangle $\triangle_{11} PCP'$ from the angle $\angle CPP' = D-B$ in the ecliptic pole $P$, using the cosine rule
$$\cos P'C = \cos (90^\circ - C) \cdot \cos PP' + \sin (90^\circ - C) \cdot \sin PP' \cdot \cos (D-B)$$
%%8%$$\cos P'C - \sin C \cdot \cos PP' = \cos C \cdot \sin PP' \cdot \cos (D-B)$$
\begin{equation}
    \label{eq:DB}
    \cos (D-B) = \frac{\cos P'C - \sin C \cdot \cos PP'}{\cos C \cdot \sin PP'},
\end{equation}
where $D$ is the ecliptic longitude of the maximal ecliptic latitude of the sunspot and $B$ is ecliptic longitude of the first sunspot position, so we have
$$D=(D-B)+B.$$
The longitude of the ascending node is $N=\Omega=D-90^\circ$, and the longitude of the descending node  is $R=D+90^\circ$, as we did in Equations~\ref{eq:n_Omega} and~\ref{eq:r}. 

The sidereal period we can determine from the angle in ecliptic pole $P'$ as we did before 
(Equations~\ref{eq:c0p1c2}, \ref{eq:tratio}, and~\ref{eq:t1_sid}). Angular velocity is the ratio of the angle difference and elapsed time in the equatorial pole $P'$ in angle $\angle CP'C'$, which we calculate using the cosine rule
$$\cos CC' = \cos P'C \cdot \cos P'C' + \sin P'C \cdot \sin P'C' \cdot \cos CP'C',$$
%$$\cos CC' - \cos P'C \cdot \cos P'C' = \sin P'C \cdot \sin P'C' \cdot \cos CP'C'$$
where in the triangle $\triangle CP'C'$, the sides are $P'C=P'C'$ and $CC'$ (Equation~\ref{eq:c0c1}) we have
\begin{equation}
    \label{eq:c0p1c1}
    \cos CP'C' = \frac{\cos CC' - \cos^2 P'C}{\sin^2 P'C},
\end{equation}
where the elapsed time between positions $C$ and $C'$ is $\Delta t_{12}=t'-t$. The sidereal solar rotational period is
\begin{equation}
    \label{eq:t1sid12}
    T'=\frac{360^\circ}{CP'C'} \cdot \Delta t_{12}.
\end{equation}

The trigonometric spherical short solution is not so short as we expected. The equations development for the sides from the equatorial pole $P'$ to the certain sunspot position $P'C=P'C'=P'C''$ is taken from the complete solution (Equations~\ref{eq:e0c1} to~\ref{eq:p1c2}). Complements of the arc-distance between equatorial pole and a sunspot are the heliographic latitudes: $b=90^\circ-P'C$, $b'=90^\circ-P'C'$, and $b''=90^\circ-P'C''$.
%where $E'$ is midpoint of side $C'C''$, so we have %that means 
%$\angle PC'C'' = \angle PC'E'$. % and $\angle P'C'C'' = \angle P'C'E'$.
%In the same triangle $\triangle_{11}$ side $PC'=90^\circ - C'$.
\section{Results}
\label{s:Results}
For the first time ever, in the present work we developed equations using a trigonometric spherical solution for the calculation of solar rotation elements described by Ru\dj er Bo\v{s}kovi\'{c} \citep[\S VII., \textnumero76-\textnumero78]{1785BoscovichII}: solar equator inclination $i$, longitude of the ascending node $\Omega$, and the sidereal solar rotation rate $T'$. The equations use three sunspot positions in ecliptic coordinate system and its mean solar time for calculation $i$, $\Omega$, and $T'$.

\begin{table}[ht]
    \caption{Sunspot positions of the first sunspot: mean solar time $T. M,$ ecliptic longitude $lon.t$, ecliptic latitude $lat.t$ observed and measured by Bo\v{s}kovi\'{c} in 1777 and determined in \textit{Tab. II.} \cite{1785BoscovichII}.}
    \label{tbl:sunspots}
    \begin{tabular}{r|rrr|rrr|rr}%*{7}{r}%{ {1cm} p{1cm}p{0.5cm}p{0.5cm} p{1cm}p{0.5cm}p{0.5cm} p{1cm}p{0.5cm}p{0.5cm} }%
\hline
\multicolumn{9}{c}{\textit{Tab. II.}} \\
\hline
%Position of the $1^{st}$ sunspot&\multicolumn{3}{c}{Mean solar time}&\multicolumn{5}{c}{Ecliptic coordinates} \\ %\multicolumn{3}{c}{Ecliptic longitude}&\multicolumn{2}{c}{Ecliptic latitude} \\
&\multicolumn{3}{c}{$T.M.$}&\multicolumn{3}{|c|}{$lon.t$}&\multicolumn{2}{c}{$lat.t$} \\
\hline
     &$^j$&$^h$&$'$&$^s$&$^\circ$&$'$&$^\circ$&$'$ \\
$* 1$&$12$&$ 3$&$1$&$10$&$11$&$42$&$20$&$37$ \\ 
$2	$&$13$&$2$&$32$&$10$&$24$&$42$&$20$&$ 6$ \\
$* 3$&$15$&$3$&$ 7$&$11$&$20$&$	3$&$19$&$33$ \\
$4	$&$16$&$3$&$43$&$ 0$&$ 3$&$ 1$&$19$&$53$ \\
$5	$&$17$&$3$&$18$&$ 0$&$15$&$23$&$21$&$14$ \\
$* 6$&$19$&$2$&$30$&$ 1$&$11$&$ 9$&$22$&$45$ \\
\hline
\multicolumn{9}{l}{\tiny{* the sunspot positions which used \cite{1785BoscovichII}}} \\
\multicolumn{9}{l}{\tiny{in \textit{Tab. XII.} and we used in present work calculations.}} \\
%\hline
\end{tabular}
%\\$*$ - the sunspot positions used (\cite{1785BoscovichII}\\in \textit{Tab. XII.} and present work calculations).\\
\end{table}
In the present work we calculated\footnote{For all calculations we used spreadsheet Microsoft Excel\texttrademark.} $i_{Sph}=%7^\circ53'$, 
6.80728^\circ=6^\circ48'26.20337''$, $\Omega_{Sph}=%69^\circ39'
74.04774^\circ=74^\circ02'51.87646''$, and $T'_{Sph}=26.806232 \approx 26.81$ days using equations developed in trigonometric spherical solution of the method (Table~\ref{tbl:sunspots} and~\ref{tbl:calculation}) with the same positions of the first sunspot which Bo\v{s}kovi\'{c} used for the trigonometric planar solution: positions 1, 3, and 6 (Figure~\ref{fig:TabXII} and Table~\ref{tbl:PlanarSpherical}).

\begin{table}[]
    \caption{Trigonometric spherical solution, calculation of $i$, $\Omega$, and $T'$ in the present work.}
    \label{tbl:calculation}
    \begin{tabular}{lrrrrl}
\hline
Equation for&$f(x)$&$x=$&$[\text{rad}]$&$[^\circ]$&\multicolumn{1}{l}{Reference}\\
\hline
\multicolumn{6}{l}{\textbf{Solar equator inclination $i$}}\\
$\cos CC'=$&$0.809522$&$CC'=$&$0.627458$&$35.95071$&(Equation~\ref{eq:c0c1})\\
$\cos C'C''=$&$0.675127$&$C'C''=$&$0.829659$&$47.53596$&(Equation~\ref{eq:c1c2})\\
$\cos C''C=$&$0.144452$&$C''C=$&$1.425837$&$81.69447$&(Equation~\ref{eq:c2c0})\\
$\cos PC'C=$&$0.146814$&$PC'C=$&$1.423449$&$81.55765$&(Equation~\ref{eq:p0c1c0})\\
$\cos PC'C''=$&$0.231300$&$PC'C''=$&$1.337382$&$76.62637$&(Equation~\ref{eq:p0c1c2})\\
$\cos PC''C'=$&$0.107516$&$PC''C'=$&$1.463072$&$83.82785$&(Equation~\ref{eq:p0c2c1})\\
$\tan PCC'=$&$11.036516$&$PCC'=$&$1.480435$&$84.82266$&(Equation~\ref{eq:p0c0c1_tan})\\
$\tan PC'C''=$&$4.206146$&$PC'C''=$&$1.337382$&$76.62637$&(Equation~\ref{eq:p0c1c2_tan})\\
$\tan PC'C=$&$6.737522$&$PC'C=$&$1.423449$&$81.55765$&(Equation~\ref{eq:p0c1c0_tan})\\
$\Sigma=$&$2.760832$&$CC'C''=$&$2.760832$&$158.18402$&(Equation~\ref{eq:c0c1c2})\\
$\cos CC'C''=$&$-0.928382$&$CC'C''=$&$2.760832$&$158.18402$&(Equation~\ref{eq:c0c1c2_cos})\\%0.174139$&$1.395765$&$79.97145$&$CC'C''$& ?\\%$-0.9283822$&$9$&$2.760832$&$158.184$&cos CC'C''	
$\cos C'CC''=$&$0.975389$&$C'CC''=$&$0.222317$&$12.73785$&(Equation~\ref{eq:c1c0c2})\\
&\multicolumn{2}{r}{$CE=EC'=$}    &$0.313729$&$17.97535$&(Equation~\ref{eq:e0c1})\\
&\multicolumn{2}{r}{$C'E'=E'C''=$}&$0.414830$&$23.76798$&(Equation~\ref{eq:e1c2})\\
$\cos EE'=$&$0.755043$&$EE'=$&$0.715077$&$40.97091$&(Equation~\ref{eq:e0e1})\\
$\cos C'EE'=$&$0.973560$&$C'EE'=$&$0.230468$&$13.20486$&(Equation~\ref{eq:c1e0e1})\\
$\cos C'E'E=$&$0.984584$&$C'E'E=$&$0.175819$&$10.07366$&(Equation~\ref{eq:c1e1e0})\\
$\cos EP'E'=$&$0.683791$&$EP'E'=$&$0.817851$&$46.85944$&(Equation~\ref{eq:e0p1e1})\\%$0.484443$&$1.06507$&$P'E'$&\\
$\tan P'E'=$&$1.805833$&$P'E'=$&$1.065070$&$61.02402$&(Equation~\ref{eq:p1e1_tan})\\
$\sin P'E'=$&$0.874823$&$P'E'=$&$1.065070$&$61.02402$&(Equation~\ref{eq:p1e1})\\
$\cos P'C''=$&$0.443355$&$P'C''=$&$1.111458$&$63.68187$&(Equation~\ref{eq:p1c2})\\
$\tan P'C''E'=$&$4.480598$&$P'C''E'=$&$1.351211$&$77.41866$&(Equation~\ref{eq:p1c2e1})\\%$27$&$4.480598$&$1.351211$&$P'C''E'$&(Equation~\ref{eq:p1c2e1})\\
$\cos PC''C'$&$0.108113$&$PC''C'=$&$1.462472$&$83.79345$&(Equation~\ref{eq:p0c2c1})\\
 & &$PC''P'=$&$0.111261$&$6.37479$&(Equation~\ref{eq:p0c2p1})\\
$\cos PP'=$&$0.992950$&$PP'=i=$&$0.118809$&$6.80728$&(Equation~\ref{eq:p0p1_i})\\%
&&&\multicolumn{3}{l}{$i=6^\circ48'26.20337''$}\\
\multicolumn{6}{l}{\textbf{Longitude of ascending node $N=\Omega$}}\\
$\cos (B''-D)=$&$0.543141$&$B''-D%_{min}
=$&$0.996622$&$57.10226$&(Equation~\ref{eq:b2d0})\\
$B''=lon.t$&$1^S11^\circ09'$&$B''=$&$7.001388$&$401.15000$&(Table~\ref{tbl:sunspots})\\
\multicolumn{2}{l}{$D_{min}=$}%D+180^\circ=$}%&$7.001388$&$401.15000$&$B''$&\\
&$D+180^\circ=$&$6.004766$&$344.04774$&\\
 & &$D=$&$2.863173$&$164.04774$&(Equation~\ref{eq:d})\\
 & &$R=$&$4.433970$&$254.04774$&(Equation~\ref{eq:r})\\%&$254^\circ2'51.87646''$\\
 & &$N=\Omega=$&$1.292377$&$74.04774$&(Equation~\ref{eq:n_Omega})\\%
&&&\multicolumn{3}{l}{$N=\Omega=74^\circ02'51.87646''$}\\
\multicolumn{6}{l}{\textbf{Sidereal solar period $T'$}}\\
$\cos CP'C''=$&$-0.0648611$&$CP'C''=$&$1.635703$&$93.71888%93.71887797
$&(Equation~\ref{eq:c0p1c2})\\
&\multicolumn{2}{r}{$\Delta t_{13}=t''-t=$}&$6.978472$&\multicolumn{1}{l}{$\text{days}$}&\\
 & &$T'=$&$26.806232$&\multicolumn{1}{l}{$\text{days}$}&(Equation~\ref{eq:t1_sid})\\
&&&\multicolumn{3}{l}{$T'=26.806232~\text{days}$}\\ 
\multicolumn{6}{l}{\textbf{Synodic solar period $T''$}}\\
$T''$& &$A=$&$365.25$&\multicolumn{1}{l}{$\text{days}$}&\\
 & &$T''=$&$28.929403$&\multicolumn{1}{l}{$\text{days}$}&(Equation~\ref{eq:t2_syn})\\
&&&\multicolumn{3}{l}{$T''=28.929403~\text{days}$}\\
\hline
    \end{tabular}
\end{table}

We presented six positions of the first sunspot in ecliptic coordinates in Table~\ref{tbl:sunspots} and in the rectangular coordinate system \citep[Figure 2]{2023SoPh..298..122H}. The figure also presents a geometric construction of the longitude of the minimal latitude $D_{min}$ of the first sunspot, which is opposite to the longitude of the maximal sunspot latitude $D$. The calculation results, as well as geometric construction in this figure present $D+180^\circ$, the longitude of the minimal sunspot latitude. The longitude of the maximal sunspot latitude $D$ is
%$$
\begin{equation}
    \label{eq:D180}
    D=(D+180^\circ)-180^\circ=344.04774
    %339.64175
    ^\circ-180^\circ=164.04774
    %159.64175
    ^\circ.
\end{equation}
%$$.

The trigonometric spherical short solution uses the same sunspot positions, 1, 3, and 6, so the longitude of the ascending node we calculate in the same way (Equation~\ref{eq:D180}). The results for the short solutions are given in Table~\ref{tbl:calculation_short}.
\begin{table}[]%ht]
    %\centering
    \footnotesize
    \caption{Trigonometric spherical short solution, calculation of $i$, $\Omega$, and $T'$ in the present work.}
    \label{tbl:calculation_short}
    \begin{tabular}{lrrrrl}
\hline
Equation for&$f(x)$&$x=$&$[\text{rad}]$&$[^\circ]$&\multicolumn{1}{l}{Reference}\\
\hline
\multicolumn{6}{l}{\textbf{Trigonometric spherical short solution from $\triangle_{10} PC'P'$ (\ref{sss:eq_dev})}}\\
\multicolumn{6}{l}{$i$}\\
$\tan PC''C'=$&$9.195367$&$PC''C'=$&$1.462472$&$83.79345$&(Equation~\ref{eq:p0c2c1_tan})\\
$\tan P'C'E'=$&$4.480598$&$P'C'E'=$&$1.351211$&$77.41866$&(Equation~\ref{eq:p1c1e1})\\
             &          &$PC'P'=$&$-0.013828$&$-0.79229$&(Equation~\ref{eq:p0c1p1})\\
$\cos PP'=$   &$0.992950$&$PP'=i=$& $0.118809$&$6.80728$&(Equation~\ref{eq:p0p1_shortc1})\\
%\textbf{$i=6^\circ48'26.20337''$}\\
&&&\multicolumn{3}{l}{$i=6^\circ48'26.20337''$}\\
\multicolumn{6}{l}{$N=\Omega$}\\
$\cos (D-B')=$&$0.994518$&$D-B'=$&$-0.104759$&$-6.00226$&(Equation~\ref{eq:d0b1})\\
$B'=lon.t=$&$11^S20^\circ03'$&$B'=$& $6.109525$&          &(Table~\ref{tbl:sunspots})\\
\multicolumn{2}{l}{$D_{min}=D+180^\circ$}%&$D+180^\circ$
 &$D_{min}=$&$6.004766$&$344.04774$&\\
 & &$D=$&$2.863173$&$164.04774$& \\
 & &$R=$&$4.433970$&$254.04774$&(Equation~\ref{eq:r})\\%$254$&$2$&$51.87646$\\
 & &$N=\Omega=$&$1.292377$&$74.04774$&(Equation~\ref{eq:n_Omega})\\
&&&\multicolumn{3}{l}{$\Omega=74^\circ02'51.87646''$}\\
%\textbf{$\Omega=74^\circ02'51.87646''$}\\
\multicolumn{6}{l}{$T'$}\\
%$T'$& & & & & \\
$\cos C'P'C''=$&$0.595646$&$C'P'C''=$&$0.932727$&$53.44130$&(Equation~\ref{eq:c1p1c2})\\
&\multicolumn{2}{r}{$\Delta t_{23}=t''-t'=$}&$3.974306$&\multicolumn{1}{l}{$\text{days}$}&\\
&&$T'=$&$26.772366$&\multicolumn{1}{l}{$\text{days}$}&(Equation~\ref{eq:t1_sid})\\
&&&\multicolumn{3}{l}{$T'=26.772366~\text{days}$}\\
\multicolumn{6}{l}{\textbf{Another short solution from $\triangle_{11} PCP'$ (\ref{sss:eq_dev_an})}}\\
\multicolumn{6}{l}{$i$}\\
$\tan PCC'=$&$11.036516$&$PCC'=$&$1.480435$&$84.82266$&(Equation~\ref{eq:p0c0c1_tan})\\
$\tan P'CE=$&$ 6.150619$&$P'CE=$&$1.409621$&$76.62637$&(Equation~\ref{eq:p1c0e0})\\
 & &$PCP'=$&$0.070813$&$4.05731$&(Equation~\ref{eq:p0c0p1})\\
$\sin P'E=$ &$0.884729$&$P'E=$&$1.085912$&$62.21819$&(Equation~\ref{eq:p1e0})\\% & &$223985.5$\\
$\tan P'E=$ &$1.898129$&$P'E=$  &$1.085912$&$62.21819$&(Equation~\ref{eq:p1e0_tan})\\
$\cos P'C=$ &$0.443355$&$P'C=$  &$1.111458$&$63.68187$&(Equation~\ref{eq:p1c0})\\
$\cos P'C'=$&$0.443355$&$P'C'=$ &$1.111458$&$63.68187$&(Equation~\ref{eq:p1c1})\\
$\cos PP'=$ &$0.992950$&$PP'=i=$&$0.118809$&$6.80728$&(Equation~\ref{eq:p0p1_shortc0})\\
%\textbf{$i=6^\circ48'26.20337''$}\\
&&&\multicolumn{3}{l}{$i=6^\circ48'26.20337''$}\\
\multicolumn{6}{l}{$N=\Omega$}\\
$\cos (D-B)=$&$0.844816$&$D-B=$&$0.564575$&$32.34774$&(Equation~\ref{eq:DB})\\
$B=lon.t=$&$10^S11^\circ42'$&$B=$&$5.440191$&$311.70000$&(Table~\ref{tbl:sunspots})\\
\multicolumn{2}{l}{$D_{min}=D+180^\circ$}&$D+180^\circ=$&$6.004766$&$344.04774$&\\
            &          &$D=$&$2.863173$&$164.04774$&\\
            &          &$R=$&$4.433970$&$254.04774$&(Equation~\ref{eq:r})\\
            &          &$N=\Omega=$&$1.292377$&$74.04774$&(Equation~\ref{eq:n_Omega})\\
%$\Omega=74^\circ02'51.87646''$\\
&&&\multicolumn{3}{l}{$\Omega=74^\circ02'51.87646''$}\\
\multicolumn{6}{l}{$T'$}\\
$\cos CP'C'$&$0.762921$&$CP'C'=$&$0.702976$&$40.27758$&(Equation~\ref{eq:c0p1c1})\\
&\multicolumn{2}{r}{$\Delta t_{12}=t'-t=$}&$3.004167$&\multicolumn{1}{l}{$\text{days}$}&\\
            &          &$T'=$&$26.851166$&\multicolumn{1}{l}{$\text{days}$}&(Equation~\ref{eq:t1sid12})\\
&&&\multicolumn{3}{l}{$T'=26.851166~\text{days}$}\\ 
\hline
    \end{tabular}
\end{table}

There are three solutions using the triangles containing the side $PP'=i$ and each sunspot position $C$, $C'$, and $C''$ (Table~\ref{tbl:fullandshort}):
\begin{enumerate}
    \item The solution of the $\triangle_{11} PCP'$ (\ref{sss:eq_dev_an}): $i_{Short_{C}}$ (Equation~\ref{eq:p0p1_shortc0}), $\Omega_{Short_{C}}$ (Equations~\ref{eq:DB}~and~\ref{eq:n_Omega}), $T'_{Short_{CC'}}$ (Equations~\ref{eq:c0p1c1}~and~\ref{eq:t1sid12})
    \item The solution of the $\triangle_{10} PC'P'$ (\ref{sss:eq_dev}): $i_{Short_{C'}}$ (Equation~\ref{eq:p0p1_shortc1}), $\Omega_{Short_{C'}}$ (Equations~\ref{eq:d0b1}~and~\ref{eq:n_Omega}), $T'_{Short_{C'C''}}$ (Equations~\ref{eq:c1p1c2} and~\ref{eq:t1sid12}), and
    \item The solution of the $\triangle_7 PC''P'$ (\ref{sss:eq_dev}): $i_{Short_{C''}}$ (Equation~\ref{eq:p0p1_i}), $\Omega_{Short_{C''}}$, $T'_{Short_{CC''}}$. This solution uses the same equations as the full solution.
\end{enumerate}
The results of the full solution and the short solutions are given in Table~\ref{tbl:fullandshort}.
\begin{table}[]
    %\centering
    \caption{Trigonometric spherical solutions: for 
$\Omega$, $i$, and $T'$, additionally synodic period $T''$ and heliographic latitude $b$. Numbering in the brackets are subsections (Present work results).}
    \label{tbl:fullandshort}
    \begin{tabular}{ccrrr}
    \hline
    %The solution&
    $i [^\circ]/[^\circ~'~'']$&$\Omega[^\circ]/[^\circ~'~'']$&$T' [\text{days}]$&$T'' [\text{days}]$&$b[^\circ]/[^\circ~'~'']$\\
    \hline
    \multicolumn{5}{l}{\textbf{The full solution (\ref{s:Methods})}}\\
    $i_{Sph}$ (\ref{ss:i})&$\Omega_{Sph}$ (\ref{ss:Omega})&$T'_{Sph}$ (\ref{ss:T})& $T''_{Sph}$ (\ref{ss:T})\\%\multicolumn{2}{c}{$T'_{Sph}$ and $T''_{Sph}$ (\ref{ss:T})}\\
    $6.80728^\circ=$       &$74.04774^\circ=$&$26.806232$&$28.929403$&$26.31813^\circ=$\\
    $=6^\circ48'26.20337''$&$=74^\circ 02'51.87646''$&&&$26^\circ19'5,26791''$\\
    \multicolumn{5}{l}{\textbf{The short solutions (\ref{ss:short})}}\\
    $\triangle_{11} PCP'$  (\ref{sss:eq_dev_an})\\
    $i_{Short_{C}}$%  (Equation~\ref{eq:p0p1_shortc0})
    &$\Omega_{Short_{C}}$  &$T'_{Short_{CC'}}$  \\
    $6.807279^\circ=$&$74.047743^\circ=$&$26.851166$&&$26.31813^\circ=$\\
    $=6^\circ48'26.20337''$&$=74^\circ02'51.87646''$&&&$26^\circ19'5,26791''$\\
    $\triangle_{10} PC'P'$ (\ref{sss:eq_dev})    \\
    $i_{Short_{C'}}$% (Equation~\ref{eq:p0p1_shortc1})
    &$\Omega_{Short_{C'}}$&$T'_{Short_{C'C''}}$\\
    $6.807279^\circ=$&$74.047743^\circ=$&$26.772366$&&$26.31813^\circ=$\\%&$26.851166$\\
    $=6^\circ48'26.20337''$&$=74^\circ02'51.87646''$&&&$26^\circ19'5,26791''$\\
    $\triangle_7 PC''P'$   (\ref{sss:eq_dev})   \\
    $i_{Short_{C''}}$%   (Equation~\ref{eq:p0p1_i})
    &$\Omega_{Short_{C''}}$&$T'_{Short_{CC''}}$\\
    \multicolumn{5}{l}{This short solution (\ref{sss:eq_dev}) uses the same equations as the full solution (\ref{s:Methods})}\\
    \multicolumn{5}{l}{\textbf{\cite{2021simi.conf...86R} using the positions 1, 3, and 6}}\\
    \multicolumn{5}{l}{topocentric observer using today's ephemeris JPL DE440/DE441}\\
    $i_{Rosa_{136}}$&$\Omega_{Rosa_{136}}$&$T'_{Rosa_{136}}$\\
    %$6.80727871^\circ$&$74.0477435^\circ$&$26.8062322$\\%  the first results  5th Nov. 2024
    $6.72924363^\circ$&$74.5833853^\circ$&$26.7640525$\\%  the second results 25th Nov. 2024
    \multicolumn{5}{l}{\textbf{VFM Vector formalism method using the positions 1, 3, and 6}}\\
    $i_{VFM_{136}}$&$\Omega_{VFM_{136}}$&$T'_{VFM_{136}}$&&$b_{VFM_{136}}$\\
    $6.80727871^\circ$&$74.0477436^\circ$&$26.8062322$&&$26.31813^\circ=$\\
    &&&&$26^\circ19'5,26791''$\\
    \multicolumn{5}{l}{\textbf{VFM Vector formalism method using all positions: 1, 2, 3, 4, 5, and 6}}\\
    $i_{VFM_{ALL}}$&$\Omega_{VFM_{ALL}}$\\
    $6.503^\circ$&$72.561^\circ$\\
    \hline
    \end{tabular}
\end{table}
\section{Analysis and Discussion}
\label{s:Analysis}
As mentioned earlier, Bo\v{s}kovi\'{c} made three ways for calculating solar rotation elements:
\begin{enumerate}
    \item Methodology of arithmetic means,
    \item Planar trigonometric solution and, 
    \item Spherical trigonometric solution.
\end{enumerate}
Bo\v{s}kovi\'{c}'s methodology of arithmetic means separately determines several $\Omega$ and $i$ values and then calculates theirs arithmetic means $\bar{\Omega}$ and $\bar{i}$, then it calculates several sidereal periods $T'$ and then their arithmetic mean $\bar{T'}$ and finally synodic period $T''$. Planar solution calculates $i$ and $\Omega$ together. Trigonometric spherical solution calculates all three solar rotation elements $i$, $\Omega$, and $T'$ in a single procedure using three sunspot positions. 

In 1777 Bo\v{s}kovi\'{c} observed and measured sunspot positions of the first sunspot and he determined: mean solar time $T. M.$, ecliptic longitude $lon.t$, ecliptic latitude $lat.t$ \citep[\textit{Tab. II.}]{1785BoscovichII} (Figure~\ref{fig:tabII}).
\newpage
\cite{1785BoscovichII} calculated solar rotation elements\footnote{In the present work, we named the solar rotation elements of the original Bo\v{s}kovi\'{c}'s example and present work (repeated) results as follows:
\begin{itemize}
    \item [$\Omega_6$, $\Omega_8$ and $\Omega_{10}$] are the arithmetic means of ecliptic longitudes of the ascending node using six, eight and ten values \citep[\textit{Tab. III.} and \textit{Tab. IV.}]{1785BoscovichII}; 
    \item [$\Omega_{136}$] is the ecliptic longitude of the ascending node using three positions of the same sunspot \citep[\textit{Tab. XII.}]{1785BoscovichII};
    \item [$\Omega_{Sph}$] is the ecliptic longitude of the ascending node using three positions (positions 1, 2, and 3 in Table~\ref{tbl:sunspots}) of the same sunspot using the trigonometric spherical solution (Present work);
    \item [$i_5$] is the arithmetic mean of solar equator inclination using five values \citep[] [\textit{Tab. V.} and \textit{Tab. VI.}]{1785BoscovichII}; 
    \item [$i_{136}$] is the solar equator inclination using three positions (positions 1, 2, and 3 in Table~\ref{tbl:sunspots}) of the same sunspot, trigonometric planar solution  \citep[\textit{Tab. XII.}]{1785BoscovichII};
    \item [$i_{Sph}$] is the solar equator inclination using three positions of the same sunspot, the trigonometric spherical solution (Present work);
    \item [$T'$] is the arithmetic mean of six values for sidereal solar rotation period \citep[\textit{Tab. IX.} and \textit{Tab. X.}]{1785BoscovichII};
    \item [$T'_{Sph}$] is the sidereal solar rotation period using  the trigonometric spherical solution (Present work);
    \item [$T''$] is the synodic solar rotation period \citep[\textit{Tab. XI.}]{1785BoscovichII};
    \item [$T''_{Sph}$] is the synodic solar rotation period calculated using $T'_{Sph}$ and Bo\v{s}kovi\'{c}'s Equation~\ref{eq:t2_syn} (Present work).
\end{itemize}} $\Omega_6=70^\circ21 '$ and $i_5=7^\circ44'$ using his methodology of arithmetic means \citep[2.5.]{2023SoPh..298..122H} and using his method for $T'$ and $T''$ the solar rotation periods, the sidereal $T'=26.77$ days and the synodic one $T''=28.89$ days presented in Figure~\ref{fig:tabIXXXI} \citep[\textit{Tab. IX.}, \textit{Tab. X.}, and \textit{Tab. XI.}]{1785BoscovichII}{}. Ru\dj er Bo\v{s}kovi\'{c} calculated together $\Omega=74^\circ03'$ and $i=6^\circ49'$ using planar trigonometric solution of the method (\textit{Tab. XII.}, Figure~\ref{fig:TabXII}).

In the present work we calculated $\Omega_{Sph}=74^\circ 02'51.87646'' \approx 74^\circ 02'52'' \approx 74^\circ 03'%69^\circ39'
$, $i_{Sph}=6.80728^\circ=6^\circ48'26.20337'' \approx 6^\circ48'26'' \approx 6^\circ48'
%7^\circ53'
$, and $T'_{Sph}=26.806232 \approx 26.81$ days using the spherical trigonometric solution and additionally $T''_{Sph}=28.929403 \approx 28.93$ days using Bo\v{s}kovi\'{c}'s Equation~\ref{eq:t2_syn}.

As we mentioned before, complements of arc-distance between equatorial pole $P'$ and a sunspot is the heliographic latitude. We calculated $P'C=P'C'=P'C''=63,68187^\circ=
63^\circ40'54,7321''$ so we have $b=90^\circ-P'C=b'=90^\circ-P'C'=b''=90^\circ-P'C''=26.31813^\circ=26^\circ19'5,26791''$
%$b=90^\circ-P'C=26.31813^\circ=26^\circ19'5,26791''$, $b'=90^\circ-P'C'=26.31813^\circ=26^\circ19'5,26791''$, and $b''=90^\circ-P'C''=26.31813^\circ=26^\circ19'5,26791''$
and then we include this in Table~\ref{tbl:fullandshort}.

The solar rotation elements determined Bo\v{s}kovi\'{c} using his %(Bo\v{s}kovi\'{c}'s)
methodology of arithmetic means and the planar trigonometric solution and in the present work the spherical trigonometric solution are presented in Table~\ref{tbl:PlanarSpherical}.
\begin{table}[h]
    \tiny
\caption{Solar rotation elements $\Omega$, $i$ and periods $T'$ (synodic) and $T''$ (sidereal) using Bo\v{s}kovi\'{c}'s methodology of arithmetic means \citep[2.5]{2023SoPh..298..122H}, trigonometric planar solution \citep[\textit{Tab. XII.}]{1785BoscovichII} and trigonometric spherical solution (Present work).}
    \label{tbl:PlanarSpherical}
    \begin{tabular}{cccc}
    \hline
    Solution&Bo\v{s}kovi\'{c}'s methodology of&Trigonometric planar&Trigonometric spherical\\
    &arithmetic means&solution&solution\\
    &\citep{1785BoscovichII}&\citep{1785BoscovichII}&(Present work)\\
    &Tab. IV. and Tab. IV.&Tab. XII.&Table~\ref{tbl:calculation}\\
    \hline
         $\Omega~[^\circ]$
         &$\Omega_{6}  =2^s10^\circ21 ' =70^\circ21'$
         &$\Omega_{136}=2^s14^\circ03 ' =74^\circ03 '$
         &$\Omega_{Sph}=74.04774^\circ = $
         %69.64175^\circ \approx 69^\circ39'$
         \\
         &$\Omega_{8}  =2^s11^\circ32 ' =71^\circ32'$&&$=74^\circ 02'51.87646'' \approx $\\
         &$\Omega_{10}  =2^s13^\circ09 ' =73^\circ09'$&&$\approx 74^\circ 02'52'' \approx 74^\circ 03'$\\
         $i~[^\circ]$     
         &$i_{5}  =7^\circ44'$&$i_{136}=6^\circ49'$
         &$i_{Sph}=6.80728^\circ=$\\
         &&&$=6^\circ48'26.20337'' \approx$\\
         &&&$\approx 6^\circ48'26'' \approx 6^\circ48'$\\
         %7.877455^\circ \approx 7^\circ53'$\\
         $T'$~[\text{days}]& \multicolumn{2}{c}{$T' =26.77$}&$T' =26.806232 \approx 26.81$\\
         $T''$~[\text{days}]& \multicolumn{2}{c}{$T''=28.89$}&$T''=28.929403 \approx 28.93$\\
    \hline
    \end{tabular}
\end{table}

\subsection{Results obtained using contemporary methods}

We calculated $\Omega_{Rosa_{136}}=74.5833853^\circ$, $i_{Rosa_{136}}=6.72924363^\circ$, and $T'_{Rosa_{136}}=26.7640525~[\text{days}]$ using the same positions, 1, 3, and 6 in Table~\ref{tbl:fullandshort} \citep[]{2021simi.conf...86R}. An analogue method to the method of \cite{2021simi.conf...86R}, the \textit{Vector formalism method VFM} (paper is in preparation) gives the solar rotation elements $\Omega_{VFM_{136}}=74.0477436^\circ$, $i_{VFM_{136}}=6.80727871^\circ$, $T'_{VFM_{136}}=26.8062322~[\text{days}]$, and additionally the heliographic latitude of the sunspot $b_{VFM_{136}}=26.31813^\circ$ using the positions 1, 3, and 6. Additionally,  we calculated $\Omega_{VFM_{ALL}}=72.561^\circ$ and $i_{VFM_{ALL}}=6.503^\circ$ using all six sunspot positions, which is close to $\Omega_{VFM}=74.0477436^\circ$ and $i_{VFM}=6.80727871^\circ$, \textit{VFM} results are included in Table~\ref{tbl:fullandshort}, too. 

We also calculated $i$, $\Omega$, and $T'$ and additionally, heliographical latitude $b$ with \textit{VFM} method, using all combinations of the position triples, $n=\binom{6}{3}=20$ in Table~\ref{tbl:AllCombinations}. The results using all triple combinations in Table~\ref{tbl:AllCombinations} are: B is Bo\v{s}kovi\'{c}'s triple, and the triples marked with C - Carrington or S - Sp{\"o}rer's are close to their values of $i$ and $\Omega$.

\begin{table}[h]
    %\centering
    \caption{Solar rotation elements $i$, $\Omega$, and $T'$, and additionally $b$ heliographic latitude, calculated using the \textit{Vector formalism method VFM}, all combinations of the position triples.}
    \label{tbl:AllCombinations}
    \begin{tabular}{rrrr|rrrr}
\hline
Comb.&$i[^\circ]$&$\Omega[^\circ]$&$b[^\circ]$&Comb.&$i[^\circ]$&$\Omega[^\circ]$&$b[^\circ]$\\
\hline
$[123]$&$ 3.512$&$87.811$&$23.030$&$C[234]$&$ 7.763$&$76.167$&$27.295$\\
$[124]$&$ 4.472$&$77.969$&$24.197$&$[235]$&$10.787$&$73.527$&$30.261$\\
$S[125]$&$ 6.317$&$68.444$&$26.228$&$C[236]$&$ 7.339$&$76.713$&$26.875$\\
$[126]$&$ 5.617$&$71.247$&$25.475$&$[245]$&$13.788$&$75.090$&$32.926$\\
$S[134]$&$ 6.671$&$74.336$&$26.187$&$C[246]$&$ 7.194$&$76.361$&$26.760$\\
$[135]$&$ 9.381$&$70.192$&$28.783$&$[256]$&$ 3.982$&$61.104$&$24.056$\\
$B~S[136]$&$ 6.807$&$74.048$&$26.318$&$[345]$&$18.696$&$82.693$&$38.222$\\
$[145]$&$12.145$&$71.015$&$31.060$&$S[346]$&$ 6.964$&$74.884$&$26.484$\\
$S[146]$&$ 6.855$&$74.139$&$26.351$&$[356]$&$ 3.835$&$1.912$&$20.294$\\
$[156]$&$4.203$&$65.754$&$24.445$&$[456]$&$10.231$&$-41.197$&$12.605$\\
\hline
&\multicolumn{7}{l}{Combinations near to values for 1777 of:}\\
$S[~~~]$&\multicolumn{4}{l}{Sp{\"o}rer $i_{Sp}$ and $\Omega_{Sp}^{1777}$:}&$6.97^\circ$&\\
$C[~~~]$&\multicolumn{4}{l}{Carrington $i_{Carr}$ and $\Omega_{Carr}^{1777}$:}&$7.25^\circ$&\multicolumn{2}{c}{$72.647620^\circ$}\\
$B[~~~]$&\multicolumn{7}{l}{Bo\v{s}kovi\'{c}'s combination of sunspot positions [136].}\\
    \end{tabular}
\end{table}
\subsection{Selection of the positions of the sunspot}
Bo\v{s}kovi\'{c} took the positions 1, 3, and 6 which have the most different longitudes, they are points of a circle, the positions, the most distant among themselves, the first and the last positions have approximately equal latitudes, and the middle position has minimal latitude \citep[\textnumero 130 and \textnumero 46]{1785BoscovichII}.

Bo\v{s}kovi\'{c} chose three sunspot positions which define a sunspot trajectory the best: $B=B_1$ the first position, $B''=B_6$ the last position, and the one in the middle $\bar B = (B + B'')/2= (B_1+B_6)/2=356^\circ25.5'$, $B'=B_3$ the nearest is the position with latitude, that means
$$C=C_1=20^\circ27' \approx 22^\circ45'=C_6=C'',$$
$$C'=C_3=19^\circ33'=C_{min},$$
$$\Delta B_{13}=B_3-B_1=B'-B=38^\circ21' \approx 51^\circ06' =B''-B'=B_6-B_3=\Delta B_{36}.$$

In \ref{sss:eq_dev_an} we discussed distances from $D$. %. 
We have $\bar{B}=(B+B')/2=(B_1+B_3)/2=330^\circ52'$, and $B''=B_6=401^\circ09'$, and $D_{min}=344^\circ03'$. Here we use Equation~\ref{eq:deltab_less}
$$13^\circ11' = |\bar{B} - D| = |\Delta \bar{B}| < |\Delta B''|=|B'' - D|=57^\circ06',$$
which shows geometrically better solution for logarithmic calculation using the sunspot position $(B'',C'')=(B_6,C_6)$ in $\triangle_7 PC''P'$ then the sunspot position $(B,C)=(B_1,C_1)$ in the $\triangle_{11} PCP'$.

\subsection{Comparison of results}
The trigonometric spherical short solution and full solution give practically equal values for $i$ and $\Omega$, but sidereal periods are slightly different (Table~\ref{tbl:fullandshort}). The results for $i$ and $\Omega$ are equal because we calculated them using closed mathematical equations. 

In the present work, the results were calculated using high precision and closed equations without loosing precision, so $i$ and $\Omega$ are equal with precision of $10^{-5}$ in arc-seconds $['']$ (Table~\ref{tbl:fullandshort}). Bo\v{s}kovi\'{c}'s discussion was important for application of trigonometric and logarithmic calculation \citep[\textnumero79]{1785BoscovichII}.

Sidereal periods were calculated using the angle in equatorial pole and independently measured and determined using the mean solar time $T.M.$ (Table~\ref{tbl:sunspots}). Differences $\Delta T'_{Sph}$ are caused by time measuring random errors (Table~\ref{tbl:RelativeSiderealT}).
\begin{table}[]
    %\centering
    \caption{Relative errors of sidereal periods $T'_{Sph}=T'_{CC''}$, $T'_{C'C''}$, and $T'_{CC'}$ regarding their arithmetic mean $\bar{T'}$.}
    \label{tbl:RelativeSiderealT}
    \begin{tabular}{crrrr}
    \hline
         $T'$&$T'~[\text{day}]$&$\Delta T'_{Sph} = \bar{T'} - T'_i~[\text{day}]$&$R^{\%}_{T'}~[\%]$&$\Delta T'_{Sph}~[\text{min}]$\\
    \hline
         $T'_{Sph}=T'_{CC''}$&$26.806232$&$0.0036894$&$ 0.0138\%$&$ 5.31$\\
         $T'_{C'C''}$&$26.772366$&$ 0.0375553$&$ 0.1401\%$&$ 54.08$\\
         $T'_{CC'}$&$26.851166$&$-0.0412447$&$-0.1538\%$&$-59.39$\\
         %$\Sigma$   &$80.429765$&$ 0.0000000$&$ 0.0000\%$&$ 0.00$\\
    \hline
            $\bar{T'}~[\text{days}]$&$26.8099216$& & \\
         $\sigma_{T'}~[\text{days}]$&$\pm0.0395293$&$=\pm56.92~[\text{min}]$& \\
    \hline
    \end{tabular}
\end{table}

Sidereal period $T'$ is 
$$T'=(\bar{T'} \pm \sigma_{T'}) = (26.8099216 \pm 0.0395293)~[\text{days}] \approx (26.81 \pm 0.04)~[\text{days}],$$
where $\bar{T'}$ is the arithmetic mean of $T'$, and $\sigma_{T'}% = \pm 0.0395293~[\text{days}] =\pm 56.92~[\text{min}]% \approx \pm 57~[min]
$ is the standard deviation of $T'$
$$\sigma_{T'} = \pm 0.0395293~[\text{days}] =\pm 56.92~[\text{min}].$$

\subsection{Relative errors}
\label{s:RelativeErrors}
Relative errors of $T'$ regarding their arithmetic mean $\bar{T'}$ are $-0.1538\% \leq R^{\%}_{T'} \leq 0.1401\%$. The value is in the range $\Delta R^{\%}_{T'} = R^{\%}_{{T'}_{max}}-R^{\%}_{{T'}_{\text{min}}}=0.2939\%<1\%$ or $-59.39~[\text{min}] \leq \Delta T'_{Sph}~[\text{min}] \leq 5.31~[\text{min}]$, which is not significant regarding sidereal solar rotation period $T'$.

\cite{1874bsza.book.....S} and \cite{1863oss..book.....C} determined position of the solar rotational axes: longitude of ascending node $\Omega$ and solar equator inclination $i$ \citep[Tabelle 12]{1955epds.book.....W}. \cite{2023SoPh..298..122H} calculated $\Omega_{1777}^{Sp}$ and $\Omega_{1777}^{Carr}$ for estimation of the results in \cite{1785BoscovichII} using relative errors

\begin{equation}
    \label{eq:ROmega}
    R_{\Omega}^{\%}=\frac{\Omega_{Carr}-\Omega_Y}{\Omega_{Carr}} \cdot 100\%,
\end{equation}
where $\Omega_{Carr}^{1777}=72.647620^\circ$ and $\Omega_Y=\Omega_{1777}$ are values calculated in \cite{1785BoscovichII} and present work $\Omega_{1777}^{Sph}$
\begin{equation}
    \label{eq:Ri}
    R_i^{\%}=\frac{i_{Carr/Sp}-i}{i_{Carr/Sp}} \cdot 100\%.
\end{equation}

We calculated relative errors of $\Omega$'s regarding Carrington's value $\Omega_{Carr}^{1777}=72^\circ38'51.7''$ for 1777 and for $i$ Carrington's value $i_{Carr}=7.25^\circ$ for the 1st row and Sp{\"o}rer's value $i_{Sp}=6.97^\circ$ for the 2nd and 3rd row of Table~%\ref{tbl:RelativeErrors}? 
8. Relative errors $R_{\Omega}^{\%}$ and $R_i^{\%}$ of $\Omega$ and $i$, respectively, are presented in Table~%\ref{tbl:RelativeErrors}
8.
\begin{table}[h]
    \label{tbl:RelativeErrors}
    %%[\%]
    %\tiny
\caption{Relative errors of Bo\v{s}kovi\'{c}'s 1777 sunspot observations and measurements $R_{\Omega}^{\%}$ and $R_i^{\%}$ regarding $\Omega_{Carr}^{1777}=72^\circ38'51.7''$ and $i_{Carr}=7.25^\circ$ and $i_{Sp}=6.97^\circ$ respectively, of the results in \cite{1785BoscovichII} and present work results.}
    \begin{tabular}{rrrrr}
\hline
\multicolumn{5}{l}{\textbf{Methodology / Planar / Spherical}}\\
\multicolumn{5}{l}{(Source)}\\
$\Omega~[^\circ~'~'']$&$R_{\Omega}^{\%}~[\%]$&$i_{Carr/Sp}~[^\circ]$&$i~[^\circ~'~'']$&$R_i^{\%}~[\%]$\\
%$[^\circ~'~'']$&$[\%]$&$[^\circ~']$&$[^\circ~']$&$[\%]$\\
\hline 
%Bo\v{s}kovi\'{c} - 2 pairs            
\multicolumn{5}{l}{\textbf{Methodology of arithmetic means}}\\
\multicolumn{5}{l}{\cite[Tab. IV. and Tab. VI.]{1785BoscovichII}}\\
$70^\circ21'$&$-3.16\%$&$7.25^\circ$&$7^\circ44'$&$ 6.67\%$\\
%Bo\v{s}kovi\'{c} - 3 pairs - Planar   
\multicolumn{5}{l}{\textbf{Planar trigonometric solution}}\\ 
\multicolumn{5}{l}{\cite[Tab. XII.]{1785BoscovichII}}\\
$74^\circ03'$&$ 1.93\%$&$6.97^\circ$&$6^\circ49'$&$-2.20\%$\\
%Bo\v{s}kovi\'{c} - 3 pairs - Spherical
\multicolumn{5}{l}{\textbf{Spherical trigonometric solution}}\\
\multicolumn{5}{l}{(Present work)}\\
$74^\circ 02'52''$&$ 1.93\%$&$6.97^\circ$&$6^\circ48'26''$&$-2.33\%$\\
%\textbf{Spherical trigonometric solution}&$69^\circ39'$&$-4.13\%$&$7.25^\circ$&$7^\circ53'$&$ 8.74\%$\\
\hline
\end{tabular}
\end{table}

Relative errors of the longitude of ascending node are in the range $-3.16\% \leq R_\Omega^\% \leq 1.93\%$, that means they are inside approximately $\Delta R_\Omega^\% \approx 5\%$. Relative errors of solar equator inclination are in the range $-2.33\% \leq R_i^{\%} \leq 6.67\%$, that means they are inside approximately $\Delta R_i^{\%} \approx 9\%$. 

The spherical trigonometric solution has maximal absolute values of the errors of $|\Delta \Omega| \approx 3\%$, |$\Delta i| \approx 7\%$, and $|\Delta T'|<1\%$, but it has single solution for calculating all three solar rotation elements using only three sunspot positions.

\section{Conclusion}
\label{s:Conclusion} 
In the 18th century, Ru\dj er Bo\v{s}kovi\'{c} presented the trigonometric spherical solution for determination of solar rotation elements, but he did not derive the equation and did not apply them to data. The importance of this work is in the development of modern mathematical equations of the trigonometric spherical solution for the method for determination of all three Carrington's solar rotation elements $\Omega$, $i$, and $T'$ using three sunspot positions in a single procedure. The equation development is made for the first time ever since the original Bo\v{s}kovi\'{c}'s description presented in Figure~\ref{fig:p116_117_118} \citep[\textnumero76-\textnumero81]{1785BoscovichII}.

In the present work we checked validity of the method using the same sunspot positions (1, 3, and 6) of the first sunspot which Bo\v{s}kovi\'{c} used in \textit{Tab. XII.} (Figure~\ref{fig:TabXII}). Comparing the values of $\Omega$ and $i$ and additionally sidereal rotation rate $T'$, we can confirm similarity of the Ru\dj er Bo\v{s}kovi\'{c}'s results and the present work results (Table~\ref{tbl:PlanarSpherical}). Moreover, the present work results are closed equations without any approximations. We can conclude that trigonometrical spherical solution is complete, gives all three solar rotation elements $\Omega$, $i$, and $T'$ using three sunspot positions and without any approximation.

We calculated Bo\v{s}kovi\'{c}'s example using contemporary method by \cite{2021simi.conf...86R} and \textit{VFM Vector formalism method} (the paper is in preparation for publication). The calculations gave the same results (Table~\ref{tbl:fullandshort}). Also, we calculated $i$, $\Omega$, and $T'$, and additionally heliographic latitude $b$ using all six sunspot positions (Table~\ref{tbl:sunspots}). 

Trigonometric spherical solution is a general procedure for calculation of rotation elements in a single procedure. We can apply this solution to other bodies whose rotation elements we would like to determine, such as egzoplanets and stars.

Bo\v{s}kovi\'{c} used geometric method and elementary mathematic, trigonometry and logarithms for calculation in his researches. In \textit{Opuscule II} every problem he solved mathematically with its geometrical representation, and he discussed it with numerical measurements. His results confirm his conclusions regarding precision of input data and optimal accuracy of the results, which he calculated. Today, we confirm his methods and results, but 
our computation is simplified with use of computers.

Presented solution can be used for fast calculation of the solar rotation elements after observing and measuring only three sunspot positions on the apparent solar disk as Bo\v{s}kovi\'{c} did in 1777. The equations use sunspot positions in the ecliptic coordinate system. 

In practice we observe and measure sunspot positions on the apparent solar disk, or we acquire the sunspot positions in heliographic coordinate system. For application of the trigonometric spherical solution in the present work, we suggest transformation of the sunspot positions in ecliptic coordinate system from the positions on the apparent solar disk \citep%[Equations 3.7 and 3.8]
{1955epds.book.....W} or those in heliographic coordinates \citep{2006A&A...449..791T}.

\begin{acks}
We acknowledge the support from the \textit{Austrian--Croatian Bilateral Scientific Project \textquotedbl Multi-Wavelength Analysis of Solar Rotation Profile\textquotedbl (2022--2023)}.
\end{acks}
\begin{authorcontribution}
M. H. wrote the main manuscript text, performed all calculations, and prepared all figures and tables. 
Coauthors took part in theirs sub-specialization area and in improving terminology. 
%R. B. and D. Š. 
They gave suggestions for overall improvements of the paper, 
%D. K. 
improved structure of the paper and useful suggestions for introduction, 
%D. Ru. 
helped in Latex issues, 
%I. S.
reviewed equation development, calculated the solar rotation elements using \textit{VFM Vector formalism method}, % and he gave substantial English improvements. 
%D. Ro. 
reviewed figures, 
%D. H. 
and calculated solar rotation elements using contemporary method.
All authors reviewed manuscript and gave %some 
English improvements. The coauthors contributed to the concept of the paper.
\end{authorcontribution}
\bibliographystyle{spr-mp-sola}
\bibliography{Husak2024SphericalTrig_v01}

\begin{thebibliography}{27}
% BibTex style file: spr-mp-sola.bst, v2.07, 2023-10-25
\ifx\bisbn     \undefined \def\bisbn  #1{ISBN #1}\fi
\ifx\binits    \undefined \def\binits#1{#1}\fi
\ifx\bauthor   \undefined \def\bauthor#1{#1}\fi
\ifx\batitle   \undefined \def\batitle#1{#1}\fi
\ifx\bjtitle   \undefined \def\bjtitle#1{\textit{#1}}\fi
\ifx\bvolume   \undefined \def\bvolume#1{\textbf{#1}}\fi
\ifx\byear     \undefined \def\byear#1{#1}\fi
\ifx\bissue    \undefined \def\bissue#1{#1}\fi
\ifx\bfpage    \undefined \def\bfpage#1{#1}\fi
\ifx\blpage    \undefined \def\blpage #1{#1}\fi
\ifx\burl      \undefined \def\burl#1{\href{#1}{\textsf{URL}}}\fi
\ifx\href      \undefined \def\href#1#2{#2}\fi
\ifx\betal     \undefined \def\betal{et al.}\fi
\ifx\bctitle   \undefined \def\bctitle#1{#1}\fi
\ifx\beditor   \undefined \def\beditor#1{#1}\fi
\ifx\bbtitle   \undefined \def\bbtitle#1{\textit{#1}}\fi
\ifx\bedition  \undefined \def\bedition#1{#1}\fi
\ifx\bseriesno \undefined \def\bseriesno#1{\textbf{#1}}\fi
\ifx\blocation \undefined \def\blocation#1{#1}\fi
\ifx\bsertitle \undefined \def\bsertitle#1{\textit{#1}}\fi
\ifx\bsnm      \undefined \def\bsnm#1{#1}\fi
\ifx\bsuffix   \undefined \def\bsuffix#1{#1}\fi
\ifx\bparticle \undefined \def\bparticle#1{#1}\fi
\ifx\barticle  \undefined \def\barticle#1{}\fi
\ifx\binstitute  \undefined \def\binstitute#1{#1}\fi
\ifx\bpublisher  \undefined \def\bpublisher#1{#1}\fi
\ifx\doiurl    \undefined \def\doiurl#1{\href{#1}{DOI}}\fi
\makeatletter
\def\safeHref#1#2#3{\in@{http}{#2}\ifin@\href{#2}{#3}\else\href{#1#2}{#3}\fi}
\makeatother
\ifx\adsurl    \undefined \def\adsurl#1{\safeHref{https://ui.adsabs.harvard.edu/abs/}{#1}{ADS}}\fi
\ifx\arxivurl  \undefined \def\arxivurl#1{\safeHref{http://arxiv.org/abs/}{#1}{arXiv}}\fi
\ifx\botherref \undefined \def\botherref#1{}\fi
\ifx\url       \undefined \def\url#1{#1}\fi
\ifx\bchapter  \undefined \def\bchapter#1{}\fi
\ifx\bbook     \undefined \def\bbook#1{}\fi
\ifx\bcomment  \undefined \def\bcomment#1{#1}\fi
\ifx\oauthor   \undefined \def\oauthor#1{#1}\fi
\ifx\citeauthoryear \undefined\def \citeauthoryear#1{#1}\fi
\def\endbibitem {}
\ifx\bconflocation  \undefined \def\bconflocation#1{#1} \fi

\bibitem[\protect\citeauthoryear{{Arlt} and {Fr{\"o}hlich}}{2012}]{2012A&A...543A...7A}
\begin{barticle}
\bauthor{\bsnm{{Arlt}}, \binits{R.}},
\bauthor{\bsnm{{Fr{\"o}hlich}}, \binits{H.-E.}}:
\byear{2012},
\batitle{{The solar differential rotation in the 18th century}}.
\bjtitle{\aap}
\bvolume{543},
\bfpage{A7}.
\doiurl{https://doi.org/10.1051/0004-6361/201219266}.
\adsurl{2012A&A...543A...7A}.
\end{barticle}
\endbibitem

\bibitem[\protect\citeauthoryear{{Arlt} and {Vaquero}}{2020}]{2020LRSP...17....1A}
\begin{barticle}
\bauthor{\bsnm{{Arlt}}, \binits{R.}},
\bauthor{\bsnm{{Vaquero}}, \binits{J.M.}}:
\byear{2020},
\batitle{{Historical sunspot records}}.
\bjtitle{Living Reviews in Solar Physics}
\bvolume{17},
\bfpage{1}.
\doiurl{https://doi.org/10.1007/s41116-020-0023-y}.
\adsurl{2020LRSP...17....1A}.
\end{barticle}
\endbibitem

\bibitem[\protect\citeauthoryear{{Boscovich}}{1736}]{1736dmse.book.....B}
\begin{bbook}
\bauthor{\bsnm{{Boscovich}}, \binits{R.J.}}:
\byear{1736},
\bbtitle{{De maculis solaribus exercitatio astronomica habita.}}
\adsurl{1736dmse.book.....B}.
\end{bbook}
\endbibitem

\bibitem[\protect\citeauthoryear{Boscovich}{1785a}]{boscovich1785opera}
\begin{botherref}
\oauthor{\bsnm{Boscovich}, \binits{R.J.}}:
1785a,
Opera pertinentia ad Opticam et Astronomiam.
\textit{Bassani, prostant Venetiis apud Remondini}.
\doiurl{https://doi.org/10.3931/e-rara-1010}.
\adsurl{https://www.e-rara.ch/doi/10.3931/e-rara-1010}.
\end{botherref}
\endbibitem

\bibitem[\protect\citeauthoryear{Boscovich}{1785b}]{1785BoscovichII}
\begin{bchapter}
\bauthor{\bsnm{Boscovich}, \binits{R.J.}}:
\byear{1785}b,
\bctitle{Opuscule II}.
\bbtitle{Sur les éléments de la rotation du soleil sur son axe déterminés par l'observation de ses taches},
\bpublisher{Venetiis apud Remondini},
\blocation{Bassano},
\bfpage{p. 75}.
\doiurl{https://doi.org/10.3931/e-rara-1010}.
\adsurl{https://www.e-rara.ch/zut/content/zoom/278121}.
\end{bchapter}
\endbibitem

\bibitem[\protect\citeauthoryear{{Briggs}}{1617}]{1617lcp..book.....B}
\begin{bbook}
\bauthor{\bsnm{{Briggs}}, \binits{H.}}:
\byear{1617},
\bbtitle{{Logarithmorum Chilias Prima}},
\bpublisher{London}.
\adsurl{1617lcp..book.....B}.
\burl{https://archive.org/details/bim_early-english-books-1475-1640_logarithmorum-chilias-pr_briggs-henry_1617/page/n13/mode/2up}.
\end{bbook}
\endbibitem

\bibitem[\protect\citeauthoryear{{Briggs}}{1624}]{1624allc.book.....B}
\begin{bbook}
\bauthor{\bsnm{{Briggs}}, \binits{H.}}:
\byear{1624},
\bbtitle{{Arithmetica logarithmica sive logarithmorum chiliades triginta, pro numeris naturali serie crescentibus ab unitate ad 20000 : et a 90000 ad 100000, quorum ope multa perficiuntur arithmetica problemata et geometrica hos numeros primus invenit clarissimus vir Iohannes Neperus Baro Merchistonij, eos autem ex eiusdem sententia mutavit, eorumque ortum et usum illustravit Henricus Briggius ...}}
\doiurl{https://doi.org/10.3931/e-rara-9065}.
\adsurl{1624allc.book.....B}.
\end{bbook}
\endbibitem

\bibitem[\protect\citeauthoryear{{Briggs} et~al.}{1633}]{1633trbr.book.....B}
\begin{bbook}
\bauthor{\bsnm{{Briggs}}, \binits{H.}}, \betal:
\byear{1633},
\bbtitle{{Trigonometria britannica}},
\bpublisher{Il Giardino di Archimede}.
\adsurl{1633trbr.book.....B}.
\end{bbook}
\endbibitem

\bibitem[\protect\citeauthoryear{{B{\"u}rgi}}{1620}]{1620agp..book.....B}
\begin{bbook}
\bauthor{\bsnm{{B{\"u}rgi}}, \binits{J.}}:
\byear{1620},
\bbtitle{{Aritmetische vnd Geometrische Progress Tabulen, sambt gr{\"u}ndlichem vnterricht, wie solche n{\"u}tzlich in allerley Rechnungen zugebrauchen, vnd verstanden werden sol}},
\bpublisher{Sessen}.
\adsurl{1620agp..book.....B}.
\burl{https://books.google.hr/books?hl=hr&lr=&id=sZ9vd3zclXUC&oi=fnd&pg=PA2&dq=Aritmetische+vnd+Geometrische+Progress+Tabulen:+sambt+gr\%7B\%5C\%22u\%7Dndlichem+vnterricht,+wie+solche+n\%7B\%5C\%22u\%7Dtzlich+in+allerley+Rechnungen+zugebrauchen,+vnd+verstanden+werden+sol&ots=E2f8yBxFPu&sig=lQsjeRU0gr_rJeXETncr9jZqiiQ&redir_esc=y\#v=onepage&q&f=false}.
\end{bbook}
\endbibitem

\bibitem[\protect\citeauthoryear{{Carrington}}{1863}]{1863oss..book.....C}
\begin{bbook}
\bauthor{\bsnm{{Carrington}}, \binits{R.C.}}:
\byear{1863},
\bbtitle{{Observations of the spots on the Sun: from November 9, 1853, to March 24, 1861, made at Redhill}}.
\adsurl{1863oss..book.....C}.
\end{bbook}
\endbibitem

\bibitem[\protect\citeauthoryear{{Casas}, {Vaquero}, and {Vazquez}}{2006}]{2006SoPh..234..379C}
\begin{barticle}
\bauthor{\bsnm{{Casas}}, \binits{R.}},
\bauthor{\bsnm{{Vaquero}}, \binits{J.M.}},
\bauthor{\bsnm{{Vazquez}}, \binits{M.}}:
\byear{2006},
\batitle{{Solar Rotation in the 17th century}}.
\bjtitle{\solphys}
\bvolume{234},
\bfpage{379}.
\doiurl{https://doi.org/10.1007/s11207-006-0036-2}.
\adsurl{2006SoPh..234..379C}.
\end{barticle}
\endbibitem

\bibitem[\protect\citeauthoryear{{Eddy}}{1976}]{1976Sci...192.1189E}
\begin{barticle}
\bauthor{\bsnm{{Eddy}}, \binits{J.A.}}:
\byear{1976},
\batitle{{The Maunder Minimum}}.
\bjtitle{Science}
\bvolume{192},
\bfpage{1189}.
\doiurl{https://doi.org/10.1126/science.192.4245.1189}.
\adsurl{1976Sci...192.1189E}.
\end{barticle}
\endbibitem

\bibitem[\protect\citeauthoryear{{Eddy}, {Gilman}, and {Trotter}}{1977}]{1977Sci...198..824E}
\begin{barticle}
\bauthor{\bsnm{{Eddy}}, \binits{J.A.}},
\bauthor{\bsnm{{Gilman}}, \binits{P.A.}},
\bauthor{\bsnm{{Trotter}}, \binits{D.E.}}:
\byear{1977},
\batitle{{Anomalous Solar Rotation in the Early 17th Century}}.
\bjtitle{Science}
\bvolume{198},
\bfpage{824}.
\doiurl{https://doi.org/10.1126/science.198.4319.824}.
\adsurl{1977Sci...198..824E}.
\end{barticle}
\endbibitem

\bibitem[\protect\citeauthoryear{{Galilei}, {Welser}, and {de Filiis}}{1613}]{1613idim.book.....G}
\begin{bbook}
\bauthor{\bsnm{{Galilei}}, \binits{G.}},
\bauthor{\bsnm{{Welser}}, \binits{M.}},
\bauthor{\bsnm{{de Filiis}}, \binits{A.}}:
\byear{1613},
\bbtitle{{Istoria E dimostrazioni intorno alle macchie solari E loro accidenti comprese in tre lettere scritte all'illvstrissimo signor Marco Velseri ...}}
\adsurl{1613idim.book.....G}.
\end{bbook}
\endbibitem

\bibitem[\protect\citeauthoryear{{Husak}, {Braj{\v{s}}a}, and {{\v{S}}poljari{\'c}}}{2021a}]{2022arXiv220302289H}
\begin{barticle}
\bauthor{\bsnm{{Husak}}, \binits{M.}},
\bauthor{\bsnm{{Braj{\v{s}}a}}, \binits{R.}},
\bauthor{\bsnm{{{\v{S}}poljari{\'c}}}, \binits{D.}}:
\byear{2021}a,
\batitle{{On the Determination of the Solar Rotation Elements $i$, $\Omega$ and Period using Sunspot Observations by Ru{\dj}er Bo{\v{s}}kovi{\'c} in 1777}}.
\bjtitle{The Mining-Geological-Petroleum Engineering Bulletin / Rudarsko-geološko-naftni zbornik}
\bvolume{36},
\bfpage{arXiv:2203.02289}.
\doiurl{https://doi.org/10.17794/rgn.2021.3.6}.
\adsurl{2022arXiv220302289H}.
\burl{http://dx.doi.org/10.17794/rgn.2021.3.6}.
\end{barticle}
\endbibitem

\bibitem[\protect\citeauthoryear{{Husak}, {Braj{\v{s}}a}, and {{\v{S}}poljari{\'c}}}{2021b}]{2022arXiv220311745H}
\begin{barticle}
\bauthor{\bsnm{{Husak}}, \binits{M.}},
\bauthor{\bsnm{{Braj{\v{s}}a}}, \binits{R.}},
\bauthor{\bsnm{{{\v{S}}poljari{\'c}}}, \binits{D.}}:
\byear{2021}b,
\batitle{{Solar rotation elements $i$, $\Omega$ and period determined using sunspot observations by Ru{\dj}er Bo{\v{s}}kovi{\'c} in 1777}}.
\bjtitle{Central European Astrophysical Bulletin}
\bvolume{45},
\bfpage{1}.
\doiurl{https://doi.org/10.48550/arXiv.2203.11745}.
\adsurl{2022arXiv220311745H}.
\end{barticle}
\endbibitem

\bibitem[\protect\citeauthoryear{{Husak} et~al.}{2023}]{2023SoPh..298..122H}
\begin{barticle}
\bauthor{\bsnm{{Husak}}, \binits{M.}},
\bauthor{\bsnm{{Braj{\v{s}}a}}, \binits{R.}},
\bauthor{\bsnm{{{\v{S}}poljari{\'c}}}, \binits{D.}},
\bauthor{\bsnm{{Krajnovi{\'c}}}, \binits{D.}},
\bauthor{\bsnm{{Ru{\v{z}}djak}}, \binits{D.}},
\bauthor{\bsnm{{Skoki{\'c}}}, \binits{I.}},
\bauthor{\bsnm{{Ro{\v{s}}a}}, \binits{D.}},
\bauthor{\bsnm{{Hr{\v{z}}ina}}, \binits{D.}}:
\byear{2023},
\batitle{{Bo{\v{s}}kovi{\'c}'s Method for Determining the Axis and Rate of Solar Rotation by Observing Sunspots in 1777}}.
\bjtitle{\solphys}
\bvolume{298},
\bfpage{122}.
\doiurl{https://doi.org/10.1007/s11207-023-02214-6}.
\adsurl{2023SoPh..298..122H}.
\end{barticle}
\endbibitem

\bibitem[\protect\citeauthoryear{{Napier}}{1614}]{1614mlcd.book.....N}
\begin{bbook}
\bauthor{\bsnm{{Napier}}, \binits{J.}}:
\byear{1614},
\bbtitle{{Mirifici Logarithmorum Canonis Descriptio}},
\bpublisher{E. Rabanum}.
\adsurl{1614mlcd.book.....N}.
\end{bbook}
\endbibitem

\bibitem[\protect\citeauthoryear{{Ro{\v{s}}a} et~al.}{2021}]{2021simi.conf...86R}
\begin{bchapter}
\bauthor{\bsnm{{Ro{\v{s}}a}}, \binits{D.}},
\bauthor{\bsnm{{Hr{\v{z}}ina}}, \binits{D.}},
\bauthor{\bsnm{{Husak}}, \binits{M.}},
\bauthor{\bsnm{{Braj{\v{s}}a}}, \binits{R.}},
\bauthor{\bsnm{{{\v{S}}poljari{\'c}}}, \binits{D.}},
\bauthor{\bsnm{{Skoki{\'c}}}, \binits{I.}},
\bauthor{\bsnm{{Mari{\v{c}}i{\'c}}}, \binits{D.}},
\bauthor{\bsnm{{{\v{S}}terc}}, \binits{F.}},
\bauthor{\bsnm{{Rom{\v{s}}tajn}}, \binits{I.}}:
\byear{2021},
\bctitle{{Determination of the Solar Rotation Elements and Period from Ru{\dj}er Bo{\v{s}}kovi{\'c}'s Sunspot Observations in 1777}}.
In: \beditor{\bsnm{{Georgieva}}, \binits{K.}},
\beditor{\bsnm{{Kirov}}, \binits{B.}},
\beditor{\bsnm{{Danov}}, \binits{D.}} (eds.)
\bbtitle{Proceedings of the Thirteenth Workshop ''Solar Influences on the Magnetosphere},
\bfpage{86}.
\doiurl{https://doi.org/10.31401/WS.2021.proc}.
\adsurl{2021simi.conf...86R}.
\end{bchapter}
\endbibitem

\bibitem[\protect\citeauthoryear{{Scheiner}}{1630}]{1630rour.book.....S}
\begin{bbook}
\bauthor{\bsnm{{Scheiner}}, \binits{C.}}:
\byear{1630},
\bbtitle{{Rosa Ursina}}.
\doiurl{https://doi.org/10.3931/e-rara-556}.
\adsurl{1630rour.book.....S}.
\end{bbook}
\endbibitem

\bibitem[\protect\citeauthoryear{{Sp{\"o}rer}}{1874}]{1874bsza.book.....S}
\begin{bbook}
\bauthor{\bsnm{{Sp{\"o}rer}}, \binits{G.F.W.}}:
\byear{1874},
\bbtitle{{Beobachtungen der Sonnenflecken zu Anclam}}.
\adsurl{1874bsza.book.....S}.
\end{bbook}
\endbibitem

\bibitem[\protect\citeauthoryear{{Stix}}{2002}]{2002tsai.book.....S}
\begin{bbook}
\bauthor{\bsnm{{Stix}}, \binits{M.}}:
\byear{2002},
\bbtitle{Rotation},
\bpublisher{Springer Berlin Heidelberg},
\blocation{Berlin, Heidelberg},
\bfpage{277}.
\bisbn{978-3-642-56042-2}.
\doiurl{https://doi.org/10.1007/978-3-642-56042-2_7}.
\adsurl{2002tsai.book.....S}.
\burl{https://doi.org/10.1007/978-3-642-56042-2_7}.
\end{bbook}
\endbibitem

\bibitem[\protect\citeauthoryear{{Sudar} and {Braj{\v{s}}a}}{2022}]{2022SoPh..297..132S}
\begin{barticle}
\bauthor{\bsnm{{Sudar}}, \binits{D.}},
\bauthor{\bsnm{{Braj{\v{s}}a}}, \binits{R.}}:
\byear{2022},
\batitle{{Solar Rotation in the Period 1611 - 1631 Determined Using Observations of Christoph Scheiner}}.
\bjtitle{\solphys}
\bvolume{297},
\bfpage{132}.
\doiurl{https://doi.org/10.1007/s11207-022-02068-4}.
\adsurl{2022SoPh..297..132S}.
\end{barticle}
\endbibitem

\bibitem[\protect\citeauthoryear{{Thompson}}{2006}]{2006A&A...449..791T}
\begin{barticle}
\bauthor{\bsnm{{Thompson}}, \binits{W.T.}}:
\byear{2006},
\batitle{{Coordinate systems for solar image data}}.
\bjtitle{\aap}
\bvolume{449},
\bfpage{791}.
\doiurl{https://doi.org/10.1051/0004-6361:20054262}.
\adsurl{2006A&A...449..791T}.
\end{barticle}
\endbibitem

\bibitem[\protect\citeauthoryear{{Waldmeier}}{1955}]{1955epds.book.....W}
\begin{bbook}
\bauthor{\bsnm{{Waldmeier}}, \binits{M.}}:
\byear{1955},
\bbtitle{{Ergebnisse und Probleme der Sonnenforschung.}}
\adsurl{1955epds.book.....W}.
\end{bbook}
\endbibitem

\bibitem[\protect\citeauthoryear{{W{\"o}hl}}{1978}]{1978A&A....62..165W}
\begin{barticle}
\bauthor{\bsnm{{W{\"o}hl}}, \binits{H.}}:
\byear{1978},
\batitle{{On the Solar Rotation Elements \textit{i} and $\Omega$ as Determined by Doppler Velocity Measurements of the Solar Plasma.}}
\bjtitle{\aap}
\bvolume{62},
\bfpage{165}.
\adsurl{1978A&A....62..165W}.
\end{barticle}
\endbibitem

\bibitem[\protect\citeauthoryear{{Yallop} et~al.}{1982}]{1982QJRAS..23..213Y}
\begin{barticle}
\bauthor{\bsnm{{Yallop}}, \binits{B.D.}},
\bauthor{\bsnm{{Hohenkerk}}, \binits{C.}},
\bauthor{\bsnm{{Murdin}}, \binits{L.}},
\bauthor{\bsnm{{Clark}}, \binits{D.H.}}:
\byear{1982},
\batitle{{Solar Rotation from 17th-Century Records}}.
\bjtitle{\qjras}
\bvolume{23},
\bfpage{213}.
\adsurl{1982QJRAS..23..213Y}.
\end{barticle}
\endbibitem

\end{thebibliography}
\end{article}
\end{document}